\pdfoutput=1
\documentclass[11pt,preprint]{aastex}   

\newcommand{\kn}{\rm}

\slugcomment{accepted for publication in ApJ}

\shorttitle{Magnetic Field Generation via the Kinetic Kelvin-Helmholtz Instability}
\shortauthors{Nishikawa et al.\ }

\begin{document}
\baselineskip 12pt

\title{Magnetic Field Generation in Core-Sheath Jets \\ via the Kinetic Kelvin-Helmholtz Instability}

\author{K.-I. Nishikawa\altaffilmark{1}, P.E. Hardee\altaffilmark{2}, I. Du\c{t}an\altaffilmark{3}, J. Niemiec\altaffilmark{4},
M. Medvedev\altaffilmark{5}, Y. Mizuno\altaffilmark{6}, \\
A. Meli\altaffilmark{7}, H. Sol\altaffilmark{8}, B. Zhang\altaffilmark{9}, M. Pohl\altaffilmark{10,11},
 D. H. Hartmann\altaffilmark{12}}
  
\altaffiltext{1}{Department of Physics,
University of Alabama in Huntsville, ZP12,
Huntsville, AL 35899, USA; ken-ichi.nishikawa@nasa.gov}
 
\altaffiltext{2}{Department of Physics and Astronomy, The University
of Alabama, Tuscaloosa, AL 35487, USA} 

\altaffiltext{3}{Institute of Space Science, Atomistilor 409, Bucharest-Magurele RO-077125, Romania}
  
\altaffiltext{4}{Institute of Nuclear Physics PAN, ul. Radzikowskiego
152, 31-342 Krak\'{o}w, Poland}

\altaffiltext{5}{Department of Physics and Astronomy, University of
Kansas, KS 66045, USA}

\altaffiltext{6}{Institute of Astronomy, National Tsing-Hua University, Hsinchu, Taiwan
30013, R.O.C.}

\altaffiltext{7}{Department  of Physics and Astronomy,
University of Gent, Proeftuinstraat 86 B-9000, 
Gent, Belgium}
 
\altaffiltext{8}{LUTH, Observatore de Paris-Meudon, 5 place Jules
Jansen, 92195 Meudon Cedex, France}
  
\altaffiltext{9}{Department of Physics, University of Nevada, Las
Vegas, NV 89154, USA}
    
\altaffiltext{10}{Institut fur Physik und Astronomie, Universit\"{a}t Potsdam, 14476 Potsdam-Golm, Germany}
\altaffiltext{11}{DESY, Platanenallee 6, 15738 Zeuthen, Germany}
   
\altaffiltext{12}{Department of Physics and Astronomy, Clemson  
University, Clemson, SC 29634, USA}

\begin{abstract}
\baselineskip 12pt
We have investigated  magnetic field generation in velocity shears via the kinetic Kelvin-Helmholtz instability (kKHI) using a relativistic plasma jet core and stationary plasma sheath. Our three-dimensional particle-in-cell simulations consider plasma jet cores with Lorentz factors of 1.5, 5, and 15 for both electron-proton and electron-positron plasmas. For electron-proton plasmas we find generation of strong large-scale DC currents and magnetic fields which extend over the entire shear-surface and reach thicknesses of a few tens of electron skin depths.  For electron-positron plasmas we find generation of alternating currents and magnetic fields.  Jet and sheath plasmas are accelerated across the shear surface in the strong magnetic fields generated by the kKHI. The mixing of jet and sheath plasmas generates transverse structure similar to that produced by the Weibel instability. 
\end{abstract}

\keywords{relativistic jets: kinetic Kelvin-Helmholtz instability - 
electron-ion plasmas, electron-positron plasmas, particle acceleration, magnetic field
generation - particle-in-cell}

\section{Introduction}

Relativistic jets are ubiquitous in astrophysical systems such as  Active Galactic Nuclei (AGN), gamma-ray bursts  (GRBs), and pulsars. Outflows interact with the ambient medium to produce relativistic shocks where  particles are accelerated and radiate in the shock magnetic fields. These shocks are collisionless and on the microscopic level are the result of beam-plasma instabilities: either electrostatic (e.g., two-stream or Buneman modes),  quasi-electrostatic (e.g., Bret et al.\ 2010), or electromagnetic (e.g., filamentation). Numerous particle-in-cell (PIC) simulations have been performed to investigate the microphysics of jet-driven collisionless relativistic shocks. The simulations demonstrate that in shocks propagating in unmagnetized or weakly magnetized plasmas
Weibel-type instabilities produce current filaments and associated magnetic fields which lead to particle acceleration and emission (e.g., Weibel 1959; Medvedev \& Loeb 1999; Frederiksen et al.\ 2004; Nishikawa et al.\ 2003, 2005, 2006, 2008, 2009; Hededal et al.\ 2004; Hededal \& Nishikawa 2005; Silva et al.\ 2003; Jaroschek et al.\ 2005; Chang et al.\ 2008; Dieckmann et al.\ 2008; Spitkovsky 2008a, 2008b; Martins et al.\ 2009, Sironi \& Spitkovsky 2009a; Haugb\/o lle 2011; Sironi, Spitokovsky,  \& Arons 2013). 

In addition to producing shocks, outflow interaction with an ambient medium includes velocity shears. In particular, the Kelvin-Helmholtz instability (KHI) has been investigated on the macroscopic level as a mean to generate magnetic fields in the presence of strong relativistic velocity shears in AGN and GRB jets (e.g., D'Angelo 1965;  Gruzinov 2008; Mizuno et al.\ 2007; Perucho \& Lobanov 2008; Zhang et al.\ 2009).  Recently PIC simulations have been employed to study magnetic field generation and particle acceleration in velocity shears at the microscopic level using counter-streaming setups.  {\kn Here the shear interactions are associated with the  kinetic Kelvin-Helmholtz instability (kKHI), also referred to as the electron-scale Kelvin-Helmholtz instability (ESKHI), e.g., Alves et al.\ 2012; Alves et al.\ 2014; Grismayer et al.\ 2013a,b; Liang et al.\ 2013a,b). }

Alves et al.\ (2012) found that alternating, hereafter AC, current and magnetic field modulations found in the non-relativistic regime are less noticeable in the relativistic regime because they are masked by strong and relatively steady, hereafter DC, current and associated magnetic field.  As the amplitude of the kKHI modulations grows the electrons from one flow cross the shear-surface and enter the counter-streaming flow. In their simulations the protons being heavier ($m_{\rm p}/m_{\rm e} = 1836$)  are unperturbed.  DC current sheets, which point in the direction of the proton velocity, form around the shear-surface. These DC current sheets induce a DC component in the magnetic field.  The DC magnetic field is dominant in the relativistic scenario because a higher DC current is set up by the slowing of electrons relative to the protons and also, because the growth rate of the AC dynamics is lowered by $\gamma_{\rm 0}^{3/2}$ compared to a non-relativistic case.  It is important to note here, that this DC magnetic field is not captured in MHD (e.g., Zhang et al.\ 2009) or fluid theories, because it results from intrinsically kinetic phenomena (currents not captured in single fluid MHD). Furthermore, since the DC field is stronger than the AC field, a kinetic treatment is clearly required in order to fully capture the generated field structure (Alves et al.\ 2012). {\rm The generated field structure is important because it may lead to a distinct radiation signature (e.g., Medvedev 2000; Sironi \& Spitkovsky 2009b; Martins et al.\ 2009; Frederiksen et al.\ 2010; Medvedev et al.\ 2011; Nishikawa et al.\ 2011, 2012).}

{\kn Grismayer et al.\ (2013a,b) have shown that the generation of DC magnetic fields in unmagnetized electron-ion shear flows is  associated to either initial thermal effects (warm shear flow) or the onset of cold shear flow electron-scale shear instabilities, in particular the cold kinetic Kelvin-Helmholtz instability. They have developed a kinetic model which predicts the growth and saturation of the DC field in both scenarios. Their multidimensional PIC simulations for an initial sharp shear confirmed their theoretical results and demonstrated the formation of long-lived magnetic fields that persist up to ion timescales (t $\sim$ 100's $\omega_{\rm pi}^{-1}$) along the full longitudinal extent of the shear layer, with a typical thickness of  $\sqrt{\gamma_{0}}c/\omega_{\rm pe}$, reaching a saturation strength $B_{\rm dc} \sim \sqrt{\gamma_{0}}v_{\rm 0}m_{\rm e}\omega_{\rm pe}/e$ that is set when the Larmor radius becomes comparable to the shear layer thickness (here $\omega_{\rm pe} \equiv (4 \pi n_0e^2/m_{\rm e})^{1/2}$). For smooth shear gradients the value of $B_{\rm dc}$ scales inversely with the initial shear gradient length scale while the layer thickness grows proportionally. Their results make it clear that the generated DC magnetic field will become dynamically important to development of the kKHI on ion time scales.}

Liang et al.\ (2013a,b) have studied the kinetic physics of relativistic shear flows using
electron-positron ($e^{\pm}$) plasmas. They have found efficient magnetic field generation 
and particle energization at the shear boundary, driven by streaming instabilities across the shear interface and 
sustained by the shear flow. Nonthermal, anisotropic high-energy particles are accelerated across field lines to 
produce a power-law tail, turning over just below the shear Lorentz factor. Additionally, Liang et al.\ (2013b) studied relativistic shear flows for hybrid positron-electron-ion ($e^{\pm}$- $i^{+}$) plasmas and compared the results to those for pure $e^{\pm}$ and pure electron-ion ($e^{-}$- $i^{+}$) plasmas. They have shown that kKHI in two different two-dimensional ($P$- and $T$- modes) simulations grows differently. Since they performed simulations using a two-dimensional system ($P$- mode), the transverse mode perpendicular
to the counter-streaming plasmas is not included in their simulations.  Among the three plasma types of relativistic shear flow, they have found that only the hybrid ($e^{\pm}$- $i^{+}$) plasma shear flow is able to energize the electrons to form a high-energy spectral peak plus  a hard power law tail. 

{\kn Alves et al.\ (2014) have extended the theoretical analysis and the simulations of the ESKHI to electron-ion plasmas with various density ratios across the shear surface, with a velocity shear gradient across the shear surface, and to warm as well as cold shear flows.  For counter-streaming flows they find that unequal densities lead to ``drift when the density symmetry is broken", the most rapid growth occurs for equal densities, that a velocity shear gradient slows the growth rate and,  as in Grismayer el al.\ (2013a,b), they find a persistent equipartition DC saturation magnetic field that ``persists longer than the proton time-scale".  A saturation electric field with $E_{\rm sat} \sim \sqrt{\gamma_0} c m_e \omega_{\rm pe}/e$ (here $\omega_{\rm pe} \equiv  \sqrt{n_{\rm e} e^2/ \epsilon_0m_{\rm e}}$) results in a maximum electron energy gain of $\Delta \mathcal{E}_{\max} \sim E_{\rm sat}c/(k_{\max}\Delta v) \propto \gamma_0^4 m_{\rm e} c^2$ where $\Delta v = v_{\rm e} - v_0$ is the difference between the accelerated electron speed and the flow speed and $1/k_{\max} = \sqrt{8/3} \gamma_0^{3/2}c/\omega_{\rm pe}$.}

We have performed three-dimensional PIC simulations to investigate the {\kn cold kKHI} using a relativistic jet core and stationary sheath  plasma  with electron-proton ($e^{-}$- $p^{+}$ with $m_{\rm p}/m_{\rm e} = 1836$) and electron-positron ($e^{\pm}$) compositions. We have compared the different plasma cases for three values of the jet core Lorentz factor: 1.5, 5, and 15. Our more physically realistic jet and stationary sheath setup allows for relativistic motions and provides a proper observer frame view of the shear layer structures. In Section 2 the simulation setup  and illustrative results are described and a theoretical analysis of the longitudinal kKHI dispersion relation is presented and compared with the simulations. The non-linear structure of electromagnetic fields and currents are discussed in Section 3.  In Section~4 the detailed dynamics of the particle mixing at the velocity shear surface is described. The results are summarized in Section~5, and their applications to AGN and GRBs are briefly discussed. 

\vspace{-0.75cm}
\section{kKHI Simulation and Theory}
\vspace{-0.2cm}
\subsection{Core-Sheath Jet Setup} 
\vspace{-0.2cm}
 
\begin{figure}[h!]
\vspace{0.3cm}
\epsscale{.99}
\plotone{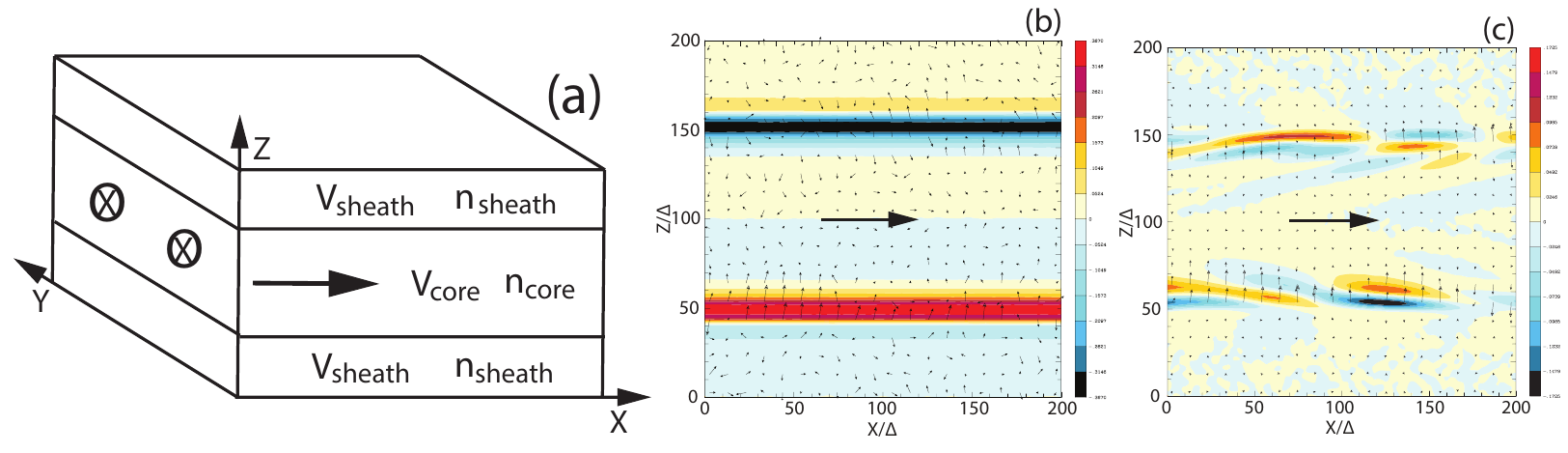}
\vspace{0.3cm}
\caption{\footnotesize \baselineskip 11.0pt  Panel (a)  shows our three-dimensional simulation setup. Panels (b) \& (c) show the magnetic field component $B_{\rm y}>0$ (red) and $B_{\rm y}<0$ (blue) plotted in the $x - z$ plane (jet flow indicated by large arrows) at the center of the simulation box, $y = 100\Delta$ at $t = 300\,\omega_{\rm pe}^{-1}$,  (b) for the $e^{-}$- $p^{+}$ case and (c) for the $e^{\pm}$ case, both with $\gamma_{\rm core}=15$.  The smaller arrows indicate the magnetic field direction in the plane.   Panels (b) \& (c) cover one fifth of the simulation system length in the $x$ direction.  The maximum and minimum magnetic field strength is $B_{\rm y} \sim$ (b) $\pm 0.367 $, and (c) $\pm  0.173$. \label{setup}}
\vspace{-0.15cm}
\end{figure}

{\kn In this simulation study we use a core-sheath plasma jet structure instead of the counter-streaming plasma setups used in previous simulations by Alves et al.\ (2012), Alves et al.\ (2014), Grismeyer et al.\ (2013a,b) and Liang et al.\ (2013a,b).} The basic setup and illustrative results are shown in Figure \ref{setup}.  In our setup a jet core with velocity $v_{\rm core}$ in the positive $x$ direction resides in the middle of the computational box.  The  upper and lower quarters of the numerical grid contain a sheath plasma  that can be stationary or moving with velocity $v_{\rm sheath}$ in the positive $x$ direction (Nishikawa et al.\ 2013a,b). 
This  model is similar to the setup in our RMHD simulations (Mizuno et al.\ 2007) that used a cylindrical jet core. However, here we represent the jet core and sheath as plasma slabs. Initially, the system is charge and current neutral. The simulations have been performed using a numerical grid with $(L_{\rm x}, L_{\rm y}, L_{\rm z}) = (1005\Delta, 205\Delta, 205\Delta)$ (simulation  cell size: $\Delta = 1$) and periodic boundary conditions in all dimensions. The jet and sheath (electron) plasma number density measured in the simulation frame is $n_{\rm jt}= n_{\rm am} = 8$.  The electron skin depth, $\lambda_{\rm s} = c/\omega_{\rm pe} = 12.2\Delta$, where $\omega_{\rm pe} = (e^{2}n_{\rm am}/\epsilon_0 m_{\rm e})^{1/2}$ is the electron plasma frequency and the electron Debye length for the ambient electrons $\lambda_{\rm D}$ is  $1.2\Delta$.  The jet-electron thermal velocity is $v_{\rm jt,th,e} = 0.014c$ in the jet reference frame, where $c$ is the speed of light.  The electron thermal velocity in the ambient plasma is $v_{\rm am,th,e} = 0.03c$, and ion thermal velocities are smaller by $(m_{\rm i}/m_{\rm e})^{1/2}$. 
Simulations were performed using an electron-positron ($e^{\pm}$) plasma or an electron-proton ($e^{-}$- $p^{+}$ with $m_{\rm p}/m_{\rm e} = 1836$) plasma for jet Lorentz factors of 1.5, 5.0, and 15.0 with the sheath plasma at rest ($v_{\rm sheath}= 0$).

An illustration of the development of the velocity shear surfaces is also shown in Figure~\ref{setup} for $e^{-}$- $p^{+}$ and $e^{\pm}$ plasmas with $v_{\rm core} = 0.9978c ~(\gamma_{\rm core}=15)$.
For the $e^{-}$- $p^{+}$ case, a nearly DC magnetic field is generated at the shear-surfaces. 
The $B_{\rm y}$ magnetic field component is generated with negative values (blue) at $z=150\Delta$ and positive values (red) at $z = 50\Delta$. Additionally, a $B_{\rm z}$ (and $B_{\rm x}$) magnetic field component, shown by the small arrows in Figure~\ref{setup}(b,c), is generated at the shear surfaces by current filaments (see  Section 3). On the other hand, for the $e^{\pm}$ case a relatively long wavelength ($\sim 100\Delta$) AC magnetic field is generated at the shear surfaces.  Note the alternating $B_{\rm y}>0$ (red) and $B_{\rm y}<0$ (blue)
in Figure \ref{setup}c along the flow direction. These results are similar to those found by Liang et al.\ (2013a,b). However, due to the two-dimensional nature of their simulations and a counter-streaming setup, there are some differences in the structure. 

\vspace{-0.75cm}
\subsection{A Longitudinal kKHI Dispersion Relation}
\vspace{-0.2cm}

{\kn We consider a sharp velocity shear surface at $z = 0$ with ``jet" plasma at $z > 0$ and ``ambient" plasma at $z < 0$ with flow in jet and/or ambient plasma in the $x$ direction.  Here the $y$-direction is infinite. Following Gruzinov (2008), Alves et al.\ (2012), Alves et al.\ (2014) and Grismayer et al.\ (2013b)} we assume uniform initial conditions on either side of the velocity shear surface,  infinitely massive ions, and perturbations to the initial conditions of the following form:
\setcounter{equation}{1}
\begin{eqnarray}
\nonumber
n(x,z,t) & = & n_0(z) + n_1(x,z,t) \\
\nonumber
{\bf v}(x,z,t) & = & v_{\rm x0}(z) + {\bf v}_1(x,z,t)\\
\eqnum{1}
{\bf E}(x,z,t) & = & E_{\rm x1}(x,z,t) + E_{\rm z1}(x,z,t)\\
\nonumber
{\bf B}(x,z,t) & = & B_{\rm y1}(x,z,t)\\
\nonumber
{\bf J}(x,z,t) & = & J_{\rm x1}(x,z,t) + J_{\rm z1}(x,z,t)
\end{eqnarray}
We extend their results to a non-counter-streaming setup.  Here we make the assumption that $v_{\rm x0} > 0$ is constant over the domain $z >0$ but can take any constant positive or negative value $v_{\rm x0} \gtrless 0$ over the domain $z<0$.  With these assumptions we look at density, velocity, current and electric field perturbations along the flow, $x$ axis, that are also a function of the normal, $z$ axis, to the velocity shear surface. The magnetic field perturbations are transverse to the flow, $y$ axis, and parallel to the shear surface. It is assumed that perturbations are of the form 
\begin{equation}
f_1(x,z,t) = f_1(z) e^{i(kx-\omega t)}~,
\end{equation}
and the wavevector ${\bf k} \equiv k_{\rm x}$ is parallel to the flow direction. Thus we are considering a velocity shear surface that is infinite transverse to the flow direction and perturbations are independent of $y$, i.e., $k_{\rm y} = 0$.

Derivation of the dispersion relation proceeds as in Alves et al.\ (2014) and the dispersion relation can be written in the following form{\footnote {See eq.(3.28) in Alves dissertation (2010).}} :
\setcounter{equation}{2}
\begin{eqnarray}
-\left[ k^2 + \omega^2_{\rm p+}/c^2 - \omega^2/c^2 \right]^{1/2}  \left[ {\omega_{\rm p+}^2/\gamma_{+}^2 \over (\omega - kv_{+})^2} - 1 \right] \left[ \left( {\omega_{\rm p+}^2 - \omega_{\rm p-}^2 \over c^2} \right) 
+   \left({\omega^2 \over c^2} - k^2 - {\omega_{\rm p+}^2 \over c^2}\right) \right]~~~~~\\
\eqnum{3}
+\left[ k^2 + \omega^2_{\rm p-}/c^2 - \omega^2/c^2 \right]^{1/2}  \left[ {\omega_{\rm p-}^2/\gamma_{-}^2 \over (\omega - kv_{-})^2} - 1 \right] \left[ \left( {\omega_{\rm p+}^2 - \omega_{\rm p-}^2 \over c^2} \right) 
-   \left({\omega^2 \over c^2} - k^2 - {\omega_{\rm p-}^2 \over c^2}\right) \right] = 0 \nonumber
\end{eqnarray}
with velocities $v_{\pm}$, associated Lorentz factors $\gamma_{\pm}$, and plasma frequencies
$
\omega_{\rm p\pm}^2 \equiv {4 \pi n_{\pm}e^2 /  \gamma_{\pm} m_{\rm e}}$ appropriate to $z_+ > 0$ and $z_- < 0$, respectively.

\vspace{-0.75cm}
\subsubsection{Analytic Solutions}
\vspace{-0.2cm}

Our generalization of previously published work to allow motion of the $z< 0$ plasma in the $\pm x$ direction, i.e., $v_{-}  \gtrless 0$, allows comparison with existing velocity shear surface counter-streaming solutions, and also allows for velocity shear surface solutions representing a high speed ``jet" plasma moving through an already relativistic ``ambient" plasma.  In particular, our generalization provides velocity shear surface solutions appropriate to spine-sheath AGN jet scenarios  (Mizuno et al.\ 2007; Hardee et al. 2007; Walg et al. 2013;  Clausen-Brown et al. 2013;  Murphy et al. 2013, references therein)  or the ``needles-in-a-jet" or ``jet-in-a-jet" scenarios proposed in the blazar AGN context (e.g., Nalewajko et al. 2011, references therein). To avoid confusion we change the notation used in eq.(3) to $n_{\rm jt} = n_+$, $n_{\rm am} = n_-$, $v_{\rm jt} = v_+$, $v_{\rm am} = v_-$, $\gamma_{\rm jt} = \gamma_+$ and $\gamma_{\rm am} = \gamma_-$. We also  use the definition
$$
 \omega_{\rm p}^2 \equiv {4 \pi n_{\rm e} e^2 \over \gamma^3 m_{\rm e}}
$$
keeping the Lorentz factor cubed in the denominator as this represents the frequency for plasma oscillations parallel to the direction of motion.  We make these changes and rewrite eq.(3) more compactly as
\setcounter{equation}{3}
\begin{eqnarray}
(k^2c^2 + \gamma_{\rm am}^2\omega_{\rm p,am}^2 - \omega^2)^{1/2}(\omega -kv_{\rm am})^2 [( \omega - kv_{\rm jt})^2 - \omega_{\rm p,jt}^2]~~~~~~~~~~~\\ 
\eqnum{4}
+ (k^2c^2 + \gamma_{\rm jt}^2\omega_{\rm p,jt}^2 - \omega^2)^{1/2}(\omega - kv_{\rm jt})^2 [( \omega - kv_{\rm am})^2 - \omega_{\rm p,am}^2] = 0~.\nonumber
\label{dsp}
\end{eqnarray}

For  counter-streaming velocities $v_{\rm am} = - v_{\rm jt} = -v_0$ and equal densities $n_{\rm jt} = n_{\rm am} = n_0$ eq.(\ref{dsp}) becomes (e.g., Gruzinov 2008)
\setcounter{equation}{4}
\begin{equation}
(k^2c^2 + \gamma_0^2\omega_{\rm p0}^2 - \omega^2)^{1/2} \{ 2(\omega^2 - k^2v_0^2)^2 - 2\omega_{\rm p0}^2(\omega^2 + k^2v_0^2) \} = 0~,
\end{equation}
with a solution  $\omega^2 = k^2c^2 + \gamma_0^2\omega_{\rm p0}^2$ that can be identified with transverse electromagnetic  waves (the electric $E_{\rm z}$ and magnetic $B_{\rm y}$ field components are transverse to the wavevector $k = k_{\rm x}$) and solutions to 
\begin{equation}
\omega^4 - (2k^2v_0^2 + \omega_{\rm p0}^2)\omega^2 + (k^2v_0^2 - \omega_{\rm p0}^2)k^2v_0^2 = 0~,
\end{equation}
given by
\begin{equation}
\omega^2 =  {{\omega_{\rm p0}^2}\over{2}} \left[ \left(1 + 2 {{k^2v_0^2}\over{\omega_{\rm p0}^2}}\right) \pm \left(1 + 8 {{k^2v_0^2}\over{\omega_{\rm p0}^2}}\right)^{1/2} \right]~,
\label{2s}
\end{equation}
that can be identified with longitudinal electrostatic plasma oscillations (the electric $E_{\rm x}$ field component is parallel to the wavevector $k_{\rm x}$).  The purely real solution, ``+" sign in eq.(\ref{2s}),  in the limit $k^2v_0^2/\omega_{\rm p0}^2 \ll 1$ is $\omega^2 \sim \omega^2_{\rm p0}$ and in the limit $k^2v_0^2/\omega_{\rm p0}^2 \gg 1$ is $\omega^2 \sim k^2v^2_0$.  The second solution, ``$-$" sign in eq.(\ref{2s}),  is purely imaginary when $k^2v_0^2/\omega_{\rm p0}^2 < 1$, has a maximum growth rate $\omega^2 = - \omega_{\rm p0}^2/8$ when $k^2v_0^2/\omega_{\rm p0}^2 =3/8$, is purely real when $k^2v_0^2/\omega_{\rm p0}^2 > 1$, and in the limit $k^2v_0^2/\omega_{\rm p0}^2 \gg 1$ becomes $\omega^2 \sim k^2v^2_0$.  This second solution is identical to the classic electrostatic two-stream instability associated with interpenetrating counter-streaming equal density relativistic plasmas. Note the difference in the ``transverse" plasma frequency  $\gamma_0^2\omega_{\rm p0}^2 = 4 \pi n_0 e^2/ \gamma_0 m_{\rm e}$ associated with transverse waves and the ``longitudinal" plasma frequency  $\omega_{\rm p0}^2 = 4 \pi n_0 e^2/ \gamma_0^3 m_{\rm e}$ associated with longitudinal waves. 
If densities in jet and ambient plasmas are unequal, $n_{\rm jt} \ne n_{\rm am}$, 
and we normalize by the ``longitudinal" plasma frequency $\omega_{\rm p,jt} = 4 \pi n_{\rm jt} e^2/ \gamma_0^3 m_{\rm e}$ and define $\omega{'} \equiv \omega / \omega_{\rm p,jt}$, $k' \equiv  kv_0/\omega_{\rm p,jt}$ and $\beta_0 \equiv v_0/c$ eq.(\ref{dsp}) reduces to eq.(29) in Alves et al.\  (2014)
\begin{eqnarray}
(\gamma_0^2 {n_{\rm am} \over n_{\rm jt}} + k'^2/\beta_0^2- \omega'^2)^{1/2} \left[ (\omega' + k')^2 - (\omega'^2 - k'^2)^2 \right]~~~~~~~~~~~\nonumber \\
+ ( \gamma_0^2 + k'^2/\beta_0^2 - \omega'^2)^{1/2} \left[ {n_{\rm am} \over n_{\rm jt}}(\omega' - k')^2 -(\omega'^2 - k'^2)^2 \right] = 0~,
\end{eqnarray}
albeit with Lorentz factors in the leading terms resulting from our ``longitudinal" as opposed to the ``transverse" plasma frequency normalization used in Alves et al.\ (2012) and Alves et al.\ (2014). Note that the transverse electromagnetic wave solution is not allowed for unequal densities on opposite sides of the velocity shear surface. 

Analytic solutions of the dispersion relation, eq.(\ref{dsp}), allowing for different densities and velocities on either side of the velocity shear surface, can be found in the low ($kc \ll \omega_{\rm p}$) and high ($kc \gg \omega_{\rm p}$) wavenumber limits. In the low wavenumber limit eq.(\ref{dsp}) can be written as
\begin{equation}
(\gamma_{\rm am}^2\omega_{\rm p,am}^2 - \omega^2)^{1/2} (\omega^2 - \omega_{\rm p,jt}^2)(\omega- kv_{\rm am})^2
+ (\gamma_{\rm jt}^2\omega_{\rm p,jt}^2 - \omega^2)^{1/2} (\omega^2 - \omega_{\rm p,am}^2)(\omega - kv_{\rm jt})^2 \sim 0~.
\label{lwl}
\end{equation}
A  complex solution to eq.(\ref{lwl}) can be written as
\begin{equation}
\omega \sim {(\gamma_{\rm am}\omega_{\rm p,jt}kv_{\rm am} + \gamma_{\rm jt}\omega_{\rm p,am}kv_{\rm jt}) \over (\gamma_{\rm am}\omega_{\rm p,jt} +  \gamma_{\rm jt}\omega_{\rm p,am})} \pm i {(\gamma_{\rm am}\omega_{\rm p,jt} \gamma_{\rm jt}\omega_{\rm p,am})^{1/2} \over (\gamma_{\rm am}\omega_{\rm p,jt} +  \gamma_{\rm jt}\omega_{\rm p,am})}k(v_{\rm jt}-v_{\rm am})~. 
\label{cmplx}
\end{equation}
{\kn In eq.(\ref{cmplx}) the real part gives the phase (drift) velocity and the imaginary part gives the temporal growth rate and directly shows the dependence of the growth rate on the velocity difference across the shear surface. Note that for counter-streaming $v_{\rm am} = -v_{\rm jt}$ the phase (drift) velocity is zero provided densities are equal on either side of the velocity shear.

In the low wavenumber limit where $v_{\rm am} = 0$ and $\gamma_{\rm am} = 1$ relevant to the numerical simulations eq.(\ref{cmplx}) becomes}
\begin{equation}
\omega \sim {(\gamma_{\rm jt}\omega_{\rm p,am}/\omega_{\rm p,jt}) \over (1 +  \gamma_{\rm jt}\omega_{\rm p,am}/\omega_{\rm p,jt})}kv_{\rm jt} \pm i {( \gamma_{\rm jt}\omega_{\rm p,am}/\omega_{\rm p,jt})^{1/2} \over (1 +  \gamma_{\rm jt}\omega_{\rm p,am}/\omega_{\rm p,jt})}kv_{\rm jt}. 
\label{lwls1}
\end{equation}
Here we see that the phase velocity (drift speed) $v_{\rm ph} \equiv \omega/k \rightarrow v_{\rm jt}$ as $\gamma_{\rm jt}\omega_{\rm p,am}/\omega_{\rm p,jt} = \gamma_{\rm jt}^{5/2} (n_{\rm am}/n_{\rm jt})^{1/2}$ increases and the temporal growth rate is maximized when $\gamma_{\rm jt}\omega_{\rm p,am}/\omega_{\rm p,jt} = \gamma_{\rm jt}^{5/2} (n_{\rm am}/n_{\rm jt})^{1/2} = 1$. {\kn In the limit where $\gamma_{\rm jt}\omega_{\rm p,am}/\omega_{\rm p,jt} =\gamma_{\rm jt}^{5/2} (n_{\rm am}/n_{\rm jt})^{1/2} \gg 1$ relevant to the numerical simulations the phase (drift) velocity is comparable to the jet speed and the low wavenumber growth rate scales with $\gamma_{\rm jt}^{-5/4}$.} The low wavenumber limit complex solution is similar in form to the hydrodynamic Kelvin-Helmholtz instability solution at low wavenumbers (Hardee 2007). In addition to the complex solution one purely real solution for $v_{\rm am} = 0$ and $\gamma_{\rm am} = 1$ and with $\gamma_{\rm jt}^2\omega_{\rm p,am}^2 > \omega_{\rm p,jt}^2$ relevant to the numerical simulations is given by 
\begin{equation}
\omega^2 \sim  \left[ {
 \gamma_{\rm jt}^2\omega_{\rm p,am}^2 - \omega_{\rm p,jt}^2 \over \omega_{\rm p,am}^2 - (2-\gamma_{\rm jt}^2)\omega_{\rm p,jt}^2}\right] \omega_{\rm p,jt}^2~,
 \label{lwls2}
\end{equation}
and in the high jet Lorentz factor limit where $\omega_{\rm p,am}^2 > \gamma_{\rm jt}^2\omega_{\rm p,jt}^2$ (recall that $\omega_{\rm p,jt}^2 \equiv 4 \pi n_{\rm jt} e^2/ \gamma^3_{\rm jt} m_{\rm e}$) becomes
$\omega^2 \sim  \gamma_{\rm jt}^2 \omega_{\rm p,jt}^2 = 4 \pi n_{\rm jt} e^2/ \gamma_{\rm jt} m_{\rm e}$.

In the high wavenumber limit eq.(\ref{dsp}) becomes
\begin{equation}
(k^2c^2 - \omega^2)^{1/2}[2(\omega -kv_{\rm am})^2 ( \omega - kv_{\rm jt})^2 - \omega_{\rm p,jt}^2
 ( \omega - kv_{\rm am})^2 - \omega_{\rm p,am}^2( \omega - kv_{\rm jt})^2] \sim 0~.
\end{equation}
Here the solution with $\omega^2 \sim k^2c^2$ for electro-magnetic waves formally exists only when the plasma frequencies on either side of the velocity shear are equal, i.e., $\gamma_{\rm am}\omega_{\rm p,am} = \gamma_{\rm jt}\omega_{\rm p,jt}$.  Two additional solutions are found from
\begin{equation}
2(\omega -kv_{\rm am})^2 ( \omega - kv_{\rm jt})^2 - \omega_{\rm p,jt}^2
 ( \omega - kv_{\rm am})^2 - \omega_{\rm p,am}^2( \omega - kv_{\rm jt})^2 \sim 0~,
\end{equation}
 where for $v_{\rm am} = 0$, as in the simulations, the solutions become 
\begin{equation}
\omega \sim kv_{\rm jt} \pm \omega_{\rm p,jt}/ \sqrt 2 ~~{\rm and}~~ \omega \sim  \pm \omega_{\rm p,am}/ \sqrt 2~,
\label{hwls}
\end{equation}
and correspond to electrostatic plasma oscillations on either side of the velocity shear surface.

\vspace{-0.75cm}
\subsubsection{Numerical Solution of the Dispersion Relation for ${\kn v_{\rm am} = 0}$}
\vspace{-0.2cm}

\begin{figure}[h!]
\vspace{0.1cm}
\epsscale{.44}
\plotone{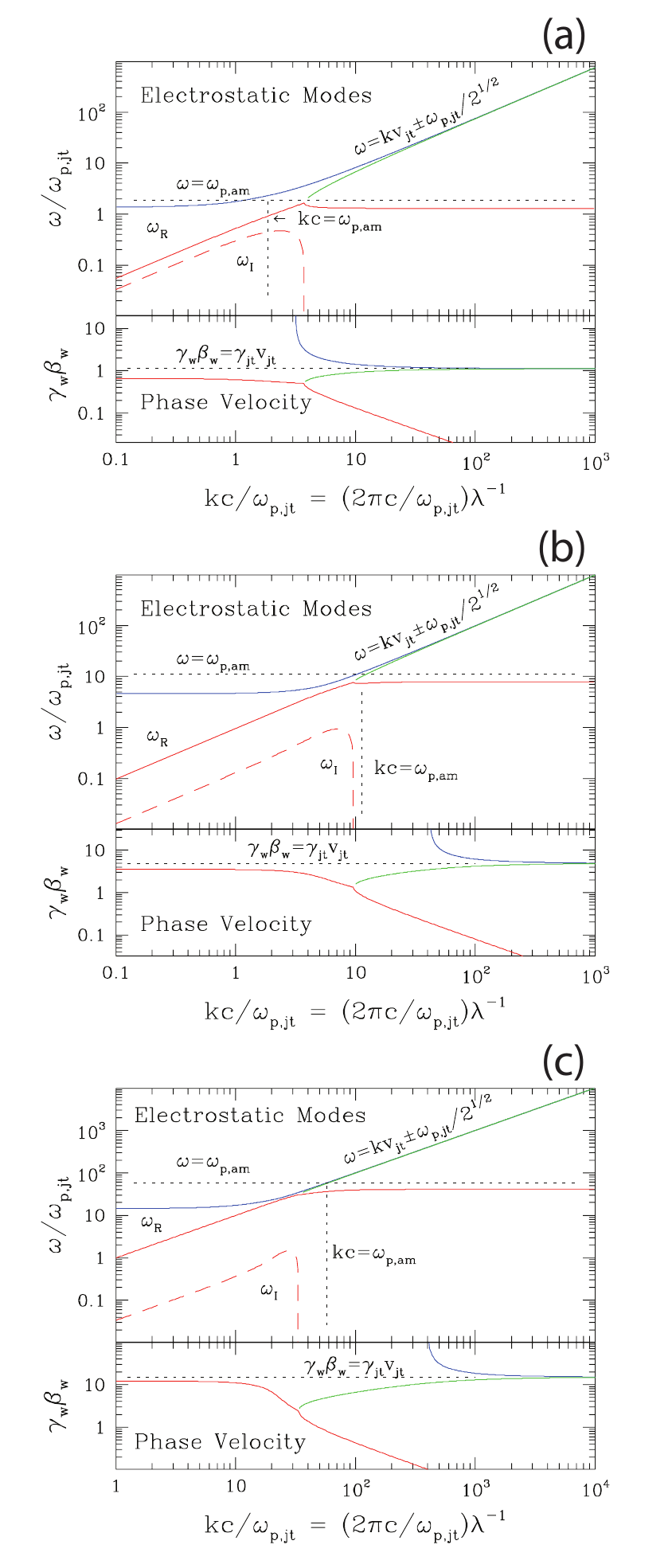}
\vspace{0.1cm}
\caption{\footnotesize \baselineskip 11pt The panels show electrostatic mode solutions, $\omega(k)$, to the dispersion relation and the phase velocity, $\gamma_w\beta_w$ where $\beta_w \equiv \omega_{\rm R}/k$, as a function of the wavenumber. The real part, $\omega_{\rm R}$, and imaginary part, $\omega_{\rm I}$, of the frequency are indicated by the solid and dashed lines, respectively, and the red, green, and blue lines indicate the different solutions. From top to bottom the panels show solutions for $\gamma_{\rm jt} =$ (a) 1.5, (b) 5.0, and (c) 15.0. 
\label{DRS}} 
\end{figure}

Numerical solution to the dispersion relation for the Lorentz factors $\gamma_{\rm jt} =$ (a) 1.5, (b) 5.0, and (c) 15.0 used in the simulations is shown in Figure \ref{DRS}.  
In all cases the jet and ambient medium are assigned equal number densities determined in the ambient (simulation) frame.  Thus, the plasma frequency ratios for the three cases  are $\omega_{\rm p,am}/\omega_{\rm p,jt} = \gamma^{3/2}_{\rm jt}  = $ (a) 1.84, (b) 11.18, and (c) 58.09.  
At small and large wavenumbers the numerical solutions agree with the analytic low, eqs.(\ref{lwls1} \& \ref{lwls2}), and high, eq.(\ref{hwls}), wavenumber solutions almost exactly.  The low wavenumber complex solution, eq.(\ref{lwls1}), provides an excellent estimate up to wavenumbers within a factor of 2 of the maximally unstable wavenumber, $k \equiv k^*$.  The large wavenumber solutions, eq.(\ref{hwls}), provide excellent estimates at wavenumbers more than a factor of 2 above the maximum marginally unstable wavenumber, $k \equiv k^{\max}$.

The numerical solutions show that the maximum growth rates are $\omega_{\rm I}^*/\omega_{\rm p,jt} =$ (a) 0.472, (b) 0.934, and (c) 1.464 at wavenumbers $k^*c/\omega_{\rm p,jt} =$ (a) 2.344, (b) 7.079, and (c) 27.542, and wavelengths $\lambda^*(\omega_{\rm p,jt}/c)  =$ (a) 2.68, (b) 0.888, and (c) 0.228. The maximum marginally unstable wavenumber is $k^{\max}c/\omega_{\rm p,jt} =$ (a) 3.715, (b) 9.550, and (c) 33.113. If we scale the growth rate and wavelength at maximum growth to the ambient plasma frequency we obtain maximum growth rates $\omega_{\rm I}^*/\omega_{\rm p,am} =$ (a) 0.256, (b) 0.079, and (c) 0.025 at wavelengths $\lambda^*  =$ (a) $4.93(c/\omega_{\rm p,am})$, (b) $9.92(c/\omega_{\rm p,am})$, and (c) $13.25 (c/\omega_{\rm p,am})$ and the minimum marginally unstable wavelength is $\lambda^{\min}  =$ (a) $3.11(c/\omega_{\rm p,am})$, (b) $7.35(c/\omega_{\rm p,am})$, and (c) $11.0 (c/\omega_{\rm p,am})$.

{\kn Numerical solution of the dispersion relation suggests that
\begin{eqnarray}
\omega_{\rm I}^* \sim  0.4 \gamma_{\rm jt}^{1/2}\omega_{\rm p,jt}~,~{\rm and}~~~~~\nonumber \\
k^*v_{\rm jt} \sim {(1 +  \gamma_{\rm jt}\omega_{\rm p,am}/\omega_{\rm p,jt})\over ( \gamma_{\rm jt}\omega_{\rm p,am}/\omega_{\rm p,jt})^{1/2}} \omega_{\rm p,jt}
\label{mxgrth}
\end{eqnarray}
provide an excellent zeroth order estimate for the maximum temporal growth rate and a reasonable zeroth order estimate for the wavenumber at maximum growth.   The maximum temporal growth rate estimate using eq.(\ref{mxgrth}) lies within 6\% of the numerically determined values.  The maximally growing wavenumber estimate using eq.(\ref{mxgrth}) ranges from (a) 20\% above to (c) 5\% below the numerically determined values where the eq.(\ref{mxgrth}) estimate has been obtained by using $\omega_{\rm I}^* \sim  \omega_{\rm p,jt}$ in  eq.(\ref{lwls1}). 

It is important to note that the maximum temporal growth rate $\omega^*_{\rm I} \propto \gamma_{\rm jt}^{-1}$ and does not decrease as rapidly with Lorentz factor as the equal density counter-streaming maximum growth rate for which $\omega^{\max}_{\rm I} \propto \gamma_{\rm jt}^{-3/2}$. It should be noted that this analysis assumes  that the ion mass is infinite and thus may not be applicable to  electron-positron cases.}

\vspace{-0.75cm}
\subsection{Longitudinal Simulation Structure}
\vspace{-0.2cm}

\begin{figure}[!h]
\vspace{-0.3cm}
\epsscale{1.0}
\plotone{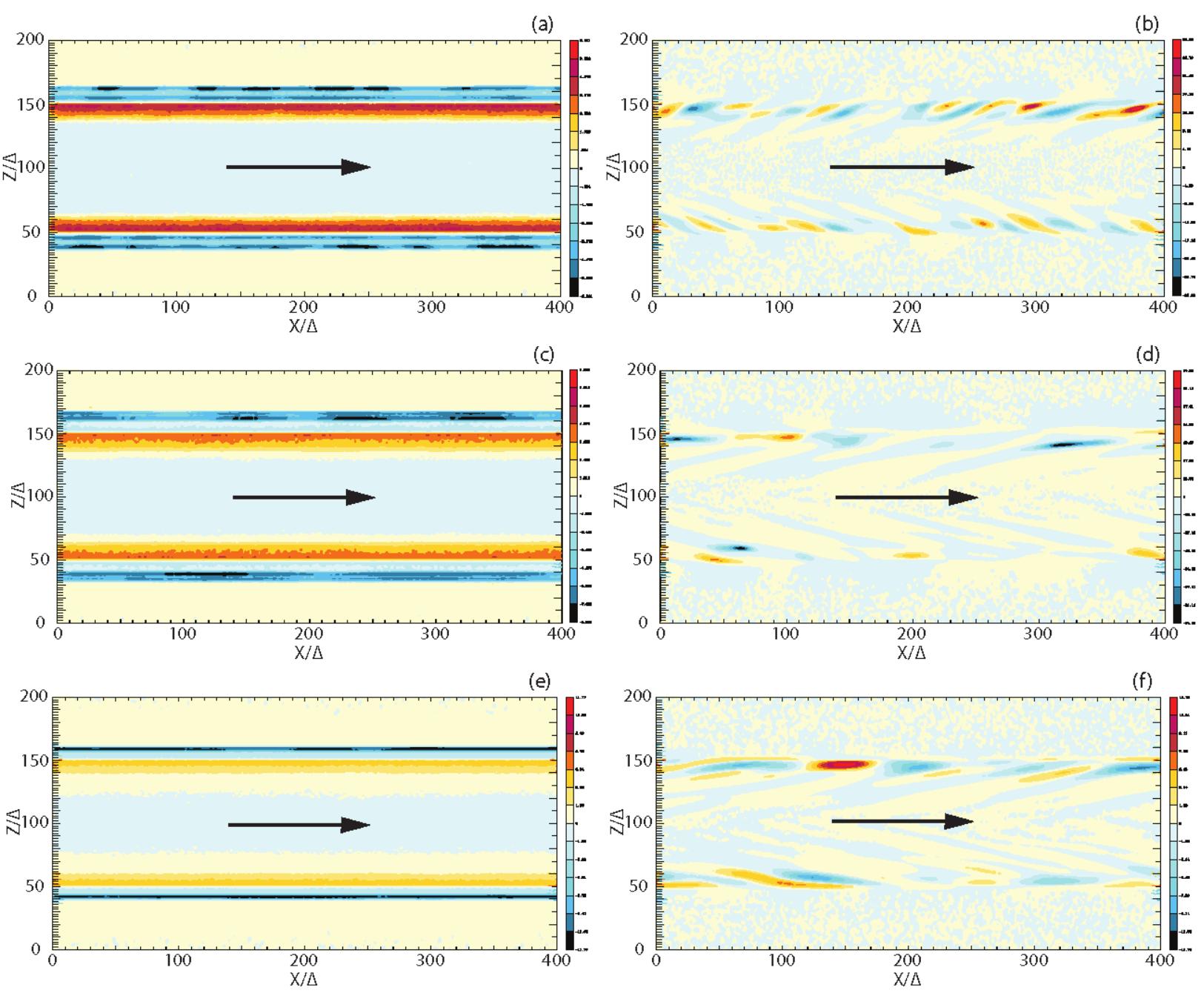}
\vspace{-0.6cm}
\caption{\footnotesize \baselineskip 11pt The panels show the $x$ component of the current density, $J_{\rm x}$, in the $x-z$ plane at $y = 100\Delta$ for  $e^-$- $p^+$ cases (left column) and  $e^{\pm}$ cases (right column) at (a \& b) $t = 200 \omega_{\rm pe}^{-1}$,
(c \& d) $t = 250 \omega_{\rm pe}^{-1}$, (e \& f) $t = 300 \omega_{\rm pe}^{-1}$.  The panels show from top to bottom $\gamma_{\rm jt} =$ (a \& b) 1.5, (c \& d) 5.0, and (e \& f) 15.0. The entire $z$-axis $0 \le z/\Delta \le 200$  is covered but only the first portion of the $x$-axis $0 \le x/\Delta \le 400$ is shown. Jet flow is indicated by the arrow. 
\label{JXxz}}
\vspace{-0.20cm}
\end{figure}

Current densities in the flow direction, $J_{\rm x}$, associated with the two velocity shear surfaces are shown in Figure \ref{JXxz}. In the $e^-$- $p^+$ cases, $J_{\rm x}$ fluctuates in strength but not direction on either side of the velocity shear surfaces and currents run parallel to the velocity shear surface. 

Currents are negative on the ambient side and positive on the jet side of the velocity shear surfaces. The fluctuations are most easily seen in the blue-black on the ambient side of the velocity shear surfaces and
the maximum and minimum current amplitude is $J_{\rm x} \sim$ (a) $\pm 6.26$, (c) $\pm 8.53$ (e) $\pm 11.77$.  In the $e^{\pm}$ cases, oblique current filaments grow at the velocity shear boundaries, and the maximum and minimum current amplitude is $J_{\rm x} \sim$ (b) $\pm 30.1$, (d) $\pm 94.7$, and  (f) $\pm 12.8$.  The $e^{\pm}$ current fluctuations lead to much larger variation in the magnetic field component, $B_{\rm y}$, associated with the velocity shear surfaces. These panels make it clear that fluctuations have the shortest spacing for the cases with $\gamma_{\rm jt} = 1.5$, fluctuation spacing is about two times larger for the cases with $\gamma_{\rm jt} = 5$, and also about two times larger for the cases with $\gamma_{\rm jt} =15$.  In the $e^{-} - p^{+} $ cases, current filaments are found along the velocity shear in the very early stages and outside the jet at later times (see Figure \ref{FigT}d).

\begin{figure}[!h]
\vspace{-0.3cm}
\epsscale{0.91}
\plotone{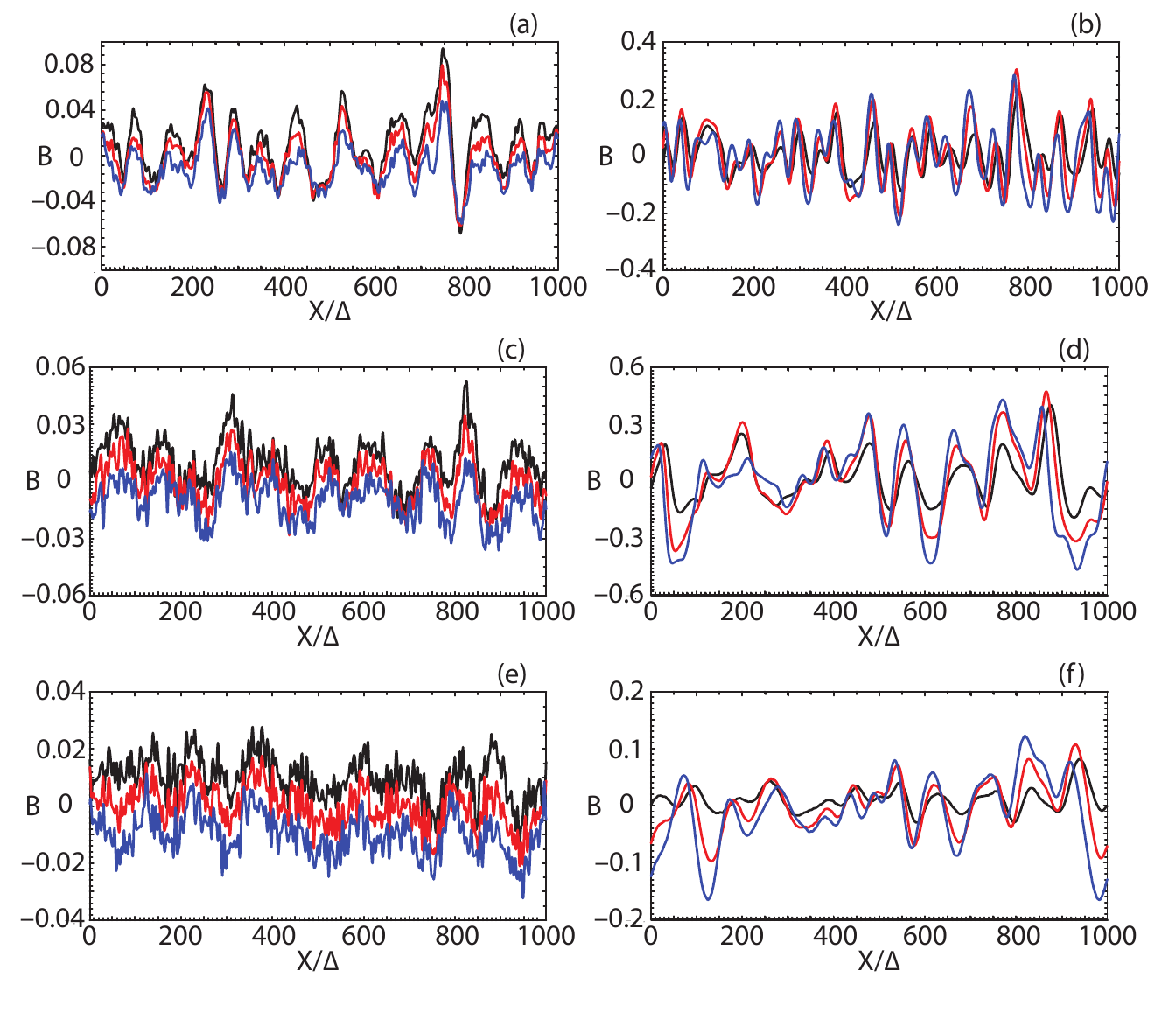}
\vspace{-0.6cm}
\caption{\footnotesize \baselineskip 11pt The panels show fluctuations in the $y$ component of the  magnetic field relative to the average, $\langle B_{\rm y} \rangle$, along 1D cuts parallel to the $x$-axis at $y = 100\Delta$ at locations $z/\Delta =$ (black line) 52, (red line) 54, (blue line) 56.  The 1D cuts are displaced vertically relative to the red line by (black  line) $+ 0.01$ and (blue line)  $- 0.01$ to separate the three lines. The panels show from top row to bottom row $\gamma_{\rm jt} =$ 1.5, 5.0, and 15.0, with the $e^-$- $p^+$ cases in the left column and the $e^{\pm}$ cases in the right column. The 1D cuts are made at the same simulation times used in Figure \ref{JXxz}, i.e., (a \& b) $t = 200 \omega_{\rm pe}^{-1}$, (c \& d) $t = 250 \omega_{\rm pe}^{-1}$, and (e \& f) $t = 300 \omega_{\rm pe}^{-1}$. 
\label{By1dZ}}
\vspace{-0.20cm}
\end{figure}

In order to make a more exact comparison with dispersion relation solutions, Figure \ref{By1dZ} shows fluctuation in 
$B_{\rm y}$ relative to the average, $\langle B_{\rm y} \rangle$, along one-dimensional cuts made parallel to the $x$-axis at $y = 100\Delta$ for $z/\Delta =$ 52, 54, \& 56.
It is important to realize that the computational box is periodic in the $x$-direction and only an integer number of wavelengths can fit in the computational box.  In the $e^-$- $p^+$ and $e^{\pm}$ $\gamma_{\rm jt} = 1.5$ cases, variation in fluctuation spacing along the $x$-axis allows $\lambda \sim 50\Delta \pm 5\Delta$, i.e., $\lambda = 1000\Delta / (20 \pm 1)$. While fluctuation amplitudes are over ten times larger for the $e^{\pm}$ case, to our measurement accuracy the $e^-$- $p^+$ and $e^{\pm}$ fluctuation wavelengths are equal.  In the $\gamma_{\rm jt} = 5$ cases we find $\lambda \sim 100\Delta \pm 10\Delta$, i.e., $\lambda = 1000\Delta / (10 \pm 1)$.  Again fluctuation amplitudes are over ten times larger for the $e^{\pm}$ case, but to our accuracy $e^-$- $p^+$ and $e^{\pm}$ wavelengths are equal.  In the $\gamma_{\rm jt} = 15$ cases we again find that $\lambda \sim 100\Delta \pm 10\Delta$ and fluctuation amplitudes are about ten times larger for the  $e^{\pm}$ case.

Comparison of the observed oscillations with the theoretically predicted fastest growing wavelength, and minimum marginally unstable wavelength associated with each Lorentz factor, suggests the following interpretation.  The observed oscillation wavelength for the $\gamma_{\rm jt} = 1.5$ cases becomes  $\lambda^{obs} \sim (4.1\pm 0.4) (c/\omega_{\rm p,am})$, recall that $c/\omega_{\rm p,am} = 12.2 \Delta$, and the observed wavelength lies between the wavelengths $\lambda^* = 4.93 (c/\omega_{\rm p,am})$ and $\lambda^{\min} = 3.11 (c/\omega_{\rm p,am})$, predicted theoretically. The observed oscillation wavelength for the $\gamma_{\rm jt} = 5$ cases becomes  $\lambda^{obs} \sim (8.2 \pm 0.8) (c/\omega_{\rm p,am})$ and also lies between the wavelengths $\lambda^* = 9.92 (c/\omega_{\rm p,am})$ and $\lambda^{\min} = 7.35 (c/\omega_{\rm p,am})$, predicted theoretically. Thus both $e^-$- $p^+$ and $e^{\pm}$ $\gamma_{\rm jt} = 1.5$ and $\gamma_{\rm jt} = 5$ cases show compelling evidence for fluctuation wavelengths near to the theoretically predicted fastest growing wavelength. Note that the predicted minimum  e-folding time of $\tau_{\rm e}^* \sim 3.9 \omega_{\rm p,am}^{-1}$ allows for $\sim 51$ e-foldings
at  $t = 200 \omega_{\rm p,am}^{-1}$ for the $\gamma_{\rm jt} = 1.5$ cases.  For the $\gamma_{\rm jt} = 5$ cases the predicted minimum  e-folding time of 
$\tau_{\rm e}^* \sim 12.7  \omega_{\rm p,am}^{-1}$ allows for  $\sim 20$ e-foldings at $t=250 \omega_{\rm p,am}^{-1}$.

For the $\gamma_{\rm jt} = 15$ cases the theoretically predicted fastest growing wavelength and minimum marginally unstable wavelength are  $\lambda^* = 13.25 (c/\omega_{\rm p,am})$ and $\lambda^{\min} = 11.0 (c/\omega_{\rm p,am})$, respectively.  The observed oscillation wavelength of $\lambda^{obs} \sim (8.2 \pm 0.8) (c/\omega_{\rm p,am})$ is somewhat shorter than the predicted minimum marginally unstable wavelength but is consistent with wave growth within the predicted unstable wavelength range. We note that the minimum e-folding time of $\tau_{\rm e}^* \sim 40 \omega_{\rm p,am}^{-1}$ has only allowed for $\sim 7$ e-foldings at $t = 300 \omega_{\rm p,am}^{-1}$ for the $\gamma_{\rm jt}=15$ cases.  This is likely insufficient time for the electrostatic mode to be fully developed and fluctuation wavelengths in these cases may be influenced by the transverse current filament structure that is discussed in Section 3. 

\vspace{-0.75cm}
\section{Field Energy Growth and Transverse Shear Surface Structure} 
\vspace{-0.2cm}
\subsection{Magnetic and Electric Field Growth}
\vspace{-0.2cm}

\begin{figure}[h!]
 \vspace{-0.5cm}
\epsscale{0.82}
\plotone{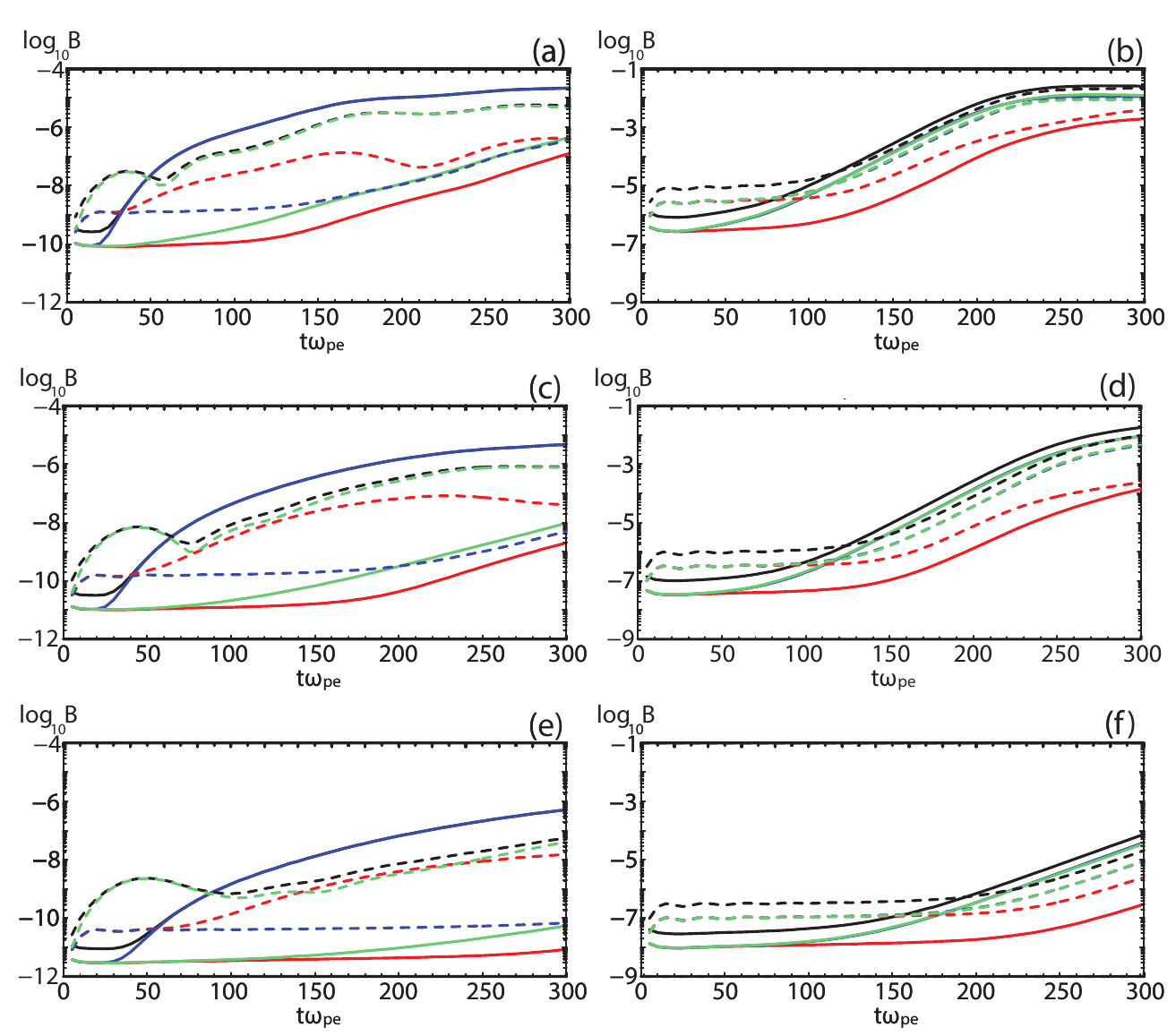}
\vspace{-0.2cm}
\caption{\footnotesize \baselineskip 11pt The panels show time evolution of the magnetic and electric field energies for  $e^-$- $p^+$  cases (left column) and $e^{\pm}$ cases  (right column) for $\gamma = 1.5$ (top row),  $\gamma = 5$ (middle row), and $\gamma = 15$ (bottom row).  Components of the magnetic field energy (solid lines) and electric field energy (dashed lines) are indicated by the red ($x$), blue ($y$) and green ($z$) lines. The black solid lines show the total magnetic field energy and the black dashed lines show the total electric field energy.  
\label{EMevol}}
\vspace{-0.4cm}
\end{figure}

Figure \ref{EMevol} shows the time evolution of  magnetic and electric field energy for the six simulation cases. 
In general, total field energy growth appears saturated for the $\gamma_{\rm jt} = 1.5$ cases, is still slowly growing for the $\gamma_{\rm jt} = 5$ cases, and remains more rapidly growing for the $\gamma_{\rm jt} = 15$ cases at the simulation end time.  The growth rate clearly decreases as the Lorentz factor increases but the growth time does not appear to have increased by the factor of $\sim 3$ ($\gamma_{\rm jt} = 5$) and $\sim 10$ ($\gamma_{\rm jt} = 15$) relative to the $\gamma_{\rm jt} = 1.5$ cases as suggested by the maximum growth rate found from the dispersion relation solutions. In the $e^-$- $p^+$ cases the total magnetic field energy exceeds the total electric field energy by factors of $\sim 4$ ($\gamma_{\rm jt} = 1.5$) to $\sim 10$ ($\gamma_{\rm jt} = 15$).  In the $e^{\pm}$ cases the total electric field energy is more comparable to the magnetic field energy with the total magnetic field energy exceeding the total electric field energy by only factors of $\gtrsim 1$ ($\gamma_{\rm jt} = 1.5$) to $\sim 4$ ($\gamma_{\rm jt} = 15$).  

In all cases the total magnetic field energy is primarily from the $B_{\rm y}$ magnetic field component. In the $e^-$- $p^+$ cases the total electric field energy is primarily from the $E_{\rm z}$ component and secondarily from the $E_{\rm x}$ component. 
Field energy associated with the $E_{\rm y}$ and $B_{\rm z} > B_{\rm x}$ field components is from one to three orders of magnitude  smaller. In the $e^-$- $p^+$ cases field energy associated with the  $E_{\rm z}$ component at first grows rapidly but then is overtaken by the growth of the field energy associated with the $B_{\rm y}$ component with an accompanying slower growth and lesser field energy associated with the  $E_{\rm x}$ field component. 
In the $e^{\pm}$ cases the total electric field energy is now primarily from the $E_{\rm y}$ component, secondarily from 
the $E_{\rm z}$ component, and only thirdly from the $E_{\rm x}$ component.  The electric field energy shows rapid initial growth, much more rapid than for the $e^-$- $p^+$ cases, that is eventually overtaken by the growth in the magnetic energy associated primarily with the $B_{\rm y}$ magnetic field component. Again there is not much energy associated with the $B_{\rm x}$ component, but there is now a significant amount of energy in the $B_{\rm z}$ field component relative to the $B_{\rm y}$ field component, unlike in the $e^-$- $p^+$ cases.

The longitudinal kKHI mode discussed in detail in Section 2.2 would lead to growth in the $E_{\rm x}$, $E_{\rm z}$, and $B_{\rm y}$ field components.  At least approximately this is in agreement with what we find for the $e^-$- $p^+$ cases, although the growth time does not increase as rapidly with the Lorentz factor as predicted.  On the other hand, in the $e^{\pm}$ cases we find the electric field energy dominated by the $E_{\rm y}$ component and a significant amount of magnetic field energy in the $B_{\rm z}$ component.  The fact that the growth of the total magnetic and electric field energies is not  decreasing with the Lorentz factor as rapidly  as predicted by the longitudinal dispersion relation and that,  particularly in the $e^{\pm}$ cases, significant magnetic and electric field components develop that are not described by the analysis in Section 2.2, provides evidence for additional processes operating in the velocity shear region which have not been captured by a longitudinal dispersion relation, and, in particular, the magnetic and electric fields imply the presence of growing transverse modes.

\vspace{-0.75cm}
\subsection{Transverse Magnetic and Current Structure}
\vspace{-0.2cm}

\begin{figure}[h!]
\vspace{-0.3cm}
\epsscale{0.93}
\plotone{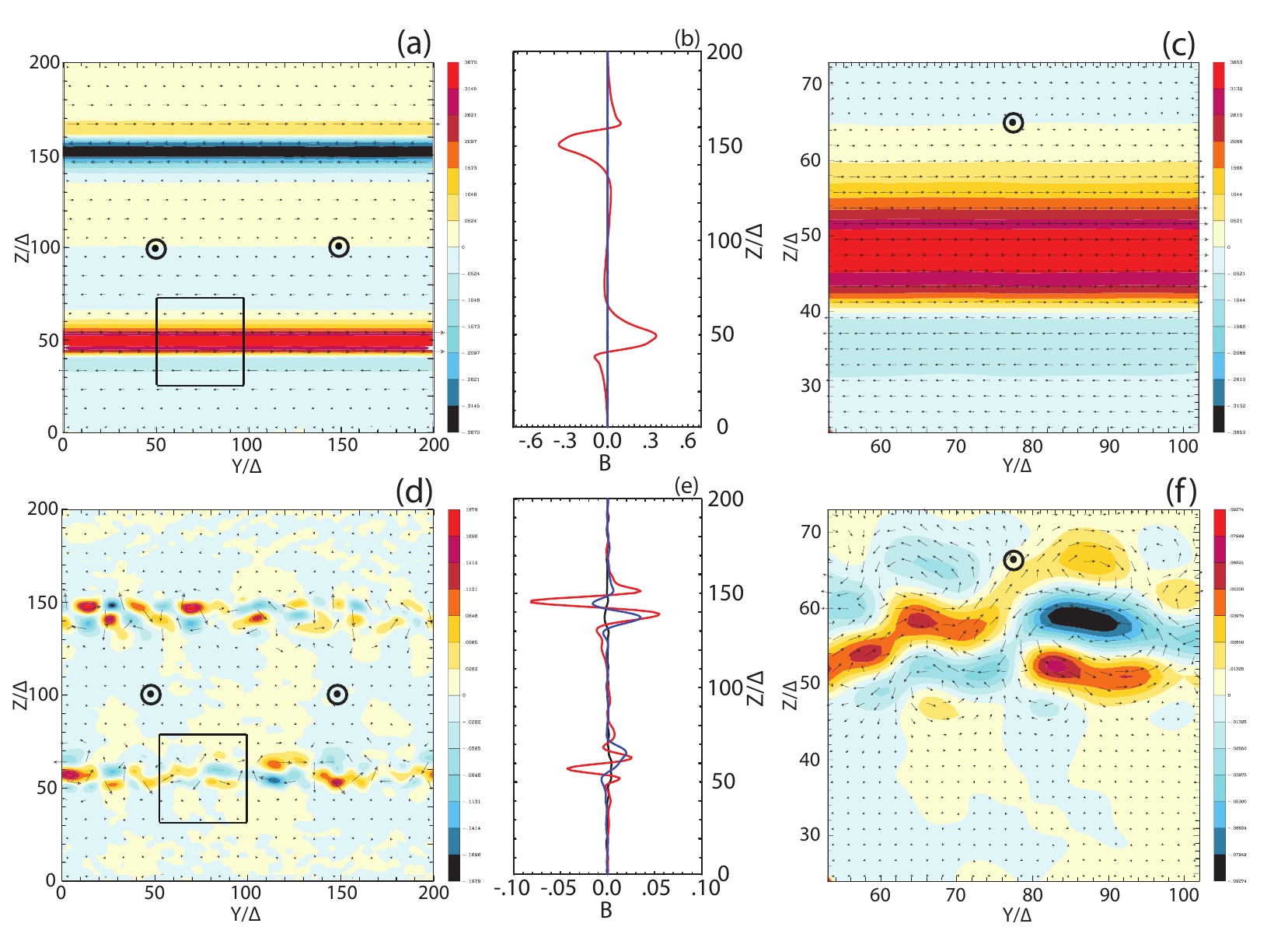}
\vspace{-0.4cm}
\caption{\footnotesize \baselineskip 11pt  Magnetic field structure transverse to the flow direction for $\gamma_{\rm jt} = 15$ is shown in the  $y - z$ plane (jet flows out of the page) at the center of the simulation box, $x = 500\Delta$ for  the  $e^-$- $p^+$  case 
(upper row) and  the $e^{\pm}$ case (lower row) at simulation time $t = 300\,\omega_{\rm pe}^{-1}$. The small arrows show the magnetic field direction in the transverse plane (the arrow length is not scaled to the magnetic field strength). 1D cuts along the $z$ axis of magnetic field components $B_{\rm x}$ (black), $B_{\rm y}$ (red), and $B_{\rm z}$ (blue) are plotted at $x = 500\Delta$ and $y = 100\Delta$ for (b) the $e^-$- $p^+$  case and (e) the $e^{\pm}$ case.  Note that the magnetic field strength scales in panels (a) ($\pm 0.367$) and (d) ($\pm 0.198$) are  different. An enlargement of the shear surface structure in the $y - z$ plane contained within the squares in the left panels is shown in the panels (c) and (f) to the right. 
\label{kkhi}}
\end{figure}

\begin{figure}[h!]
\vspace{-0.7cm}
\epsscale{.78}
\plotone{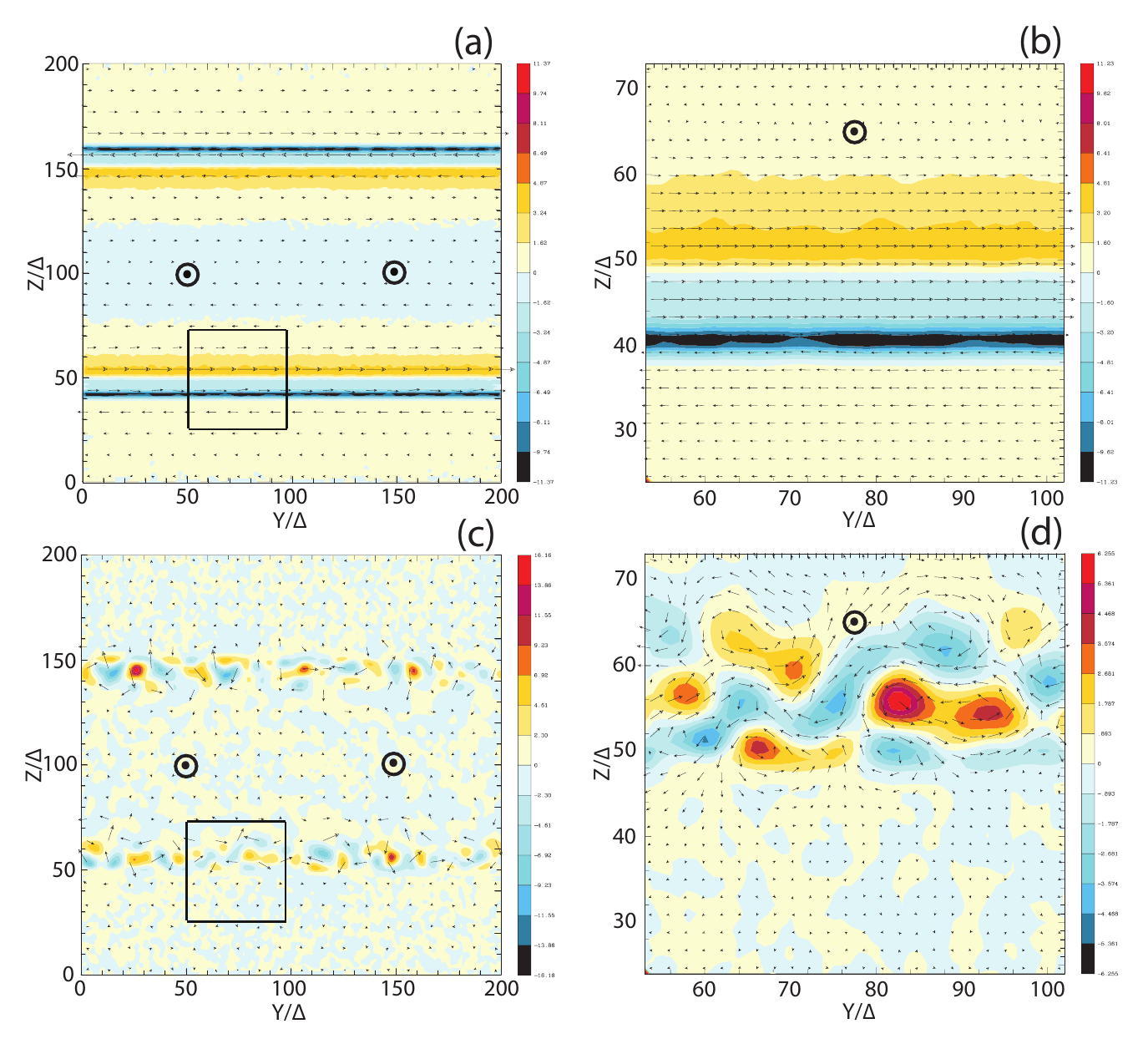}
\vspace{-0.5cm}
\caption{\footnotesize \baselineskip 11pt The  panels show the   $J_{\rm x}$  current structures   in the $y-z$ plane for panels (a \& b) the $e^-$- $p^+$ case  and for panels (c \& d) the $e^{\pm}$ case at  $t = 300 \omega_{\rm pe}^{-1}$,  both with $\gamma_{\rm jt} = 15$. The small arrows show the magnetic field direction in the transverse plane (the arrow length increases with the magnetic filed strength, but is not scaled to the magnetic field strength).  The areas within the squares in panels (a \& c)  are enlarged in panels (c \& d), respectively. The maximum and minimum of the current density is  (a) $\pm 11.37$, (b) $\pm11.23$, (c) $\pm  16.16$ and (d) $\pm6.26$. 
\label{Jxem}}
\vspace{-0.15cm}
\end{figure}

Figure \ref{kkhi} shows  the structure of the $B_{\rm y}$ component of the magnetic field in the  $y - z$ plane (jet flows out of the page) at the midpoint of the simulation box, $x = 500\Delta$, and 1D cuts along the $z$ axis showing the magnitude and direction of the magnetic field components at the midpoint of the simulation box,  $x = 500\Delta$ and $y = 100\Delta$ for the  $e^-$- $p^+$  case 
and the $e^{\pm}$ case at simulation time $t = 300\,\omega_{\rm pe}^{-1}$, both with $\gamma_{\rm jt} = 15$. 
Comparison of the transverse structures in the $y$ direction at the velocity shear surfaces shown in panels (a) and (d)  with the parallel structures in the $x$ direction shown in Figure \ref{setup} in panels (b) and (c) shows that
the fluctuations transverse to the jet in the $y$ direction are much more rapid than fluctuations along the jet in the $x$ direction. In the $e^-$- $p^+$  case, magnetic fields appear relatively uniform at the velocity shear surfaces along the transverse $y$ direction just as was seen at the velocity shear surfaces along the parallel $x$ direction, with almost no transverse fluctuations visible in the magnetic field (small fluctuations in the $y$ direction over distances on the order of $\sim 10\Delta$ are visible in the currents in Figure \ref{Jxem}b), whereas small longitudinal mode fluctuations in the $x$ direction occur over distances $\sim 100\Delta$.  For the electron-positron case, the magnetic field alternates in both the $y$ and $z$ directions and these transverse fluctuations occur over distances on the order of $\sim 10\Delta$,  whereas longitudinal mode fluctuations in the $x$ direction occur over distances $\sim 100\Delta$.  

The 1D cuts show  that the $B_{\rm y}$ field component dominates in the $e^-$- $p^+$  case, that the $B_{\rm y}$ field component is about an order of magnitude  smaller for the $e^{\pm}$ case, and that the $B_{\rm z}$ component is significant for the $e^{\pm}$ case, as already indicated in Figure \ref{EMevol}. The 1D cuts also show that there is magnetic field sign reversal on either side of the maximum that is relatively small for the $e^-$- $p^+$  case but is much more significant for the $e^{\pm}$ case, which can be seen also in Figure \ref{kkhi}d.  More details are revealed by the enlargement of the region contained in the squares. 
For the $e^-$- $p^+$ case, the generated relatively uniform DC magnetic field is symmetric about the velocity shear surface, e.g., note that $B_{\rm y} >0$ immediately around the shear surface and $B_{\rm y}<0$ in the jet and ambient plasmas at somewhat larger distances from the shear surface.  On the other hand, for the $e^{\pm}$ case the generated AC magnetic field resides largely on the jet side of the velocity shear surface.

Figure \ref{Jxem} shows how the $J_{\rm x}$ current structure in a small $y-z$ plane, responsible for the magnetic 
field structure shown in Figure \ref{kkhi}.  Motion of  electrons and/or positrons across the shear surface produces the electric currents shown also in Figure \ref{Jxem} by the arrows. Relativistic jet flow is out of the page and in the $e^-$- $p^+$ case positive (red/orange) and negative (blue/black) current flows along the jet and the sheath side of the velocity shear surfaces, respectively.  Positive currents are stronger than the negative currents, leading to the generation of the $B_{\rm y}$ magnetic field component, shown in Figure \ref{kkhi}a-c.  In the $e^{\pm}$ case, a complex current structure appears on the jet side of the velocity shear surface. The associated magnetic fields are then folded and twisted by vortical plasma motions.  The vortices appear like ``islands'' in the magnetic field. In the currents it is possible to see that the transverse fluctuation scale is similar in the $e^-$- $p^+$  and $e^{\pm}$ cases, but the structures are considerably different.

{\kn It seems likely that the development of transverse filamentary structure has influenced the longitudinal structure  studied in Section 2.  In general, we find that the kKHI grows on timescales $t \propto \gamma_{\rm jt}$, albeit growth also depends on the density ratio across the velocity shear.  Once particles have scattered across the velocity shear via kKHI or thermal motions, structure associated with interpenetrating relativistic plasmas can develop.  For $k \equiv k_{\rm x}$ there is the beam space charge instability (Bludeman, Watson \& Rosenbluth (1960) and see also  eqs.(8) \& (9) in Hardee \& Rose (1978)) with 
\begin{eqnarray}
\omega = kv_{\rm jt} \pm i {\omega_{\rm p,jt}\over \omega_{\rm p,am}} kv_{\rm jt},~{\rm and}~~~~~~~~~~~~~~~~~~~~ \nonumber \\ 
\omega_{\rm I}^{\max} = 0.69 (\omega_{\rm p,jt}^2 \omega_{\rm p,am})^{1/3}~{\rm at}~ kv_{\rm jt} = \omega_{\rm p,am} - 0.4(\omega_{\rm p,jt}^2\omega_{\rm p,am})^{1/3}~.
\end{eqnarray}
For $k \equiv k_{\rm y}$ there is the ordinary mode (filamentation) instability with
\begin{equation}
\omega_{\rm R} = 0,~{\rm and}~\omega_{\rm I} =  {{\omega_{\rm p,am}}\over {(\omega_{\rm p,am}^2 + \omega_{\rm p,jt}^2)^{1/2}}} (\gamma_{\rm jt} \omega_{\rm p,jt}) {v_{\rm jt} \over c}~.
\end{equation}
Equation (18) is formally found in the limit $k^2c^2 \gg 2\omega_{\rm p,am}^2 + \gamma_{\rm jt}^2\omega_{\rm p,jt}^2 + \omega_{\rm p,jt}^2$ and is like  eq.(11) in Hardee \& Rose (1978) which assumed $ \omega_{\rm p,am} \gg \omega_{\rm p,jt}$, but now with  $\Omega = eB/m c = 0$ and allowing for $\omega_{\rm p,jt} \sim \omega_{\rm p,am}$ to reveal the density dependence.  We have adopted the present notation in equations (17 \& 18), and note that they were derived originally in the context of electron-positron jet and ambient plasmas.  They should also apply to electron-proton plasmas where the ions are assumed infinitely massive.

We see from the above that the longitudinal beam space charge instability grows on timescales $t \propto \gamma_{\rm jt}$ that are comparable to the kKHI. On the other hand, filamentary structure associated with the transverse ordinary mode instability grows on timescales $t \propto \gamma_{\rm jt}^{1/2}$ and thus can grow faster than kKHI longitudinal structure for large Lorentz factors. While there is excellent agreement between observed longitudinal structure scales and theoretical prediction for the two lower Lorentz factors, such is not the case for the high Lorentz factor simulation. Here we believe that more rapid growth of transverse structure in the high Lorentz factor case has overwhelmed slower growth of the longitudinal kKHI and led to a longitudinal length scale that is less than the minimum unstable wavelength associated with the kKHI.}

\vspace{-0.75cm}
\subsection{Lorentz factor Differences at the Shear Surface}
\vspace{-0.2cm}

\begin{figure}[h!]
\vspace{-0.5cm}
\epsscale{0.88}
\plotone{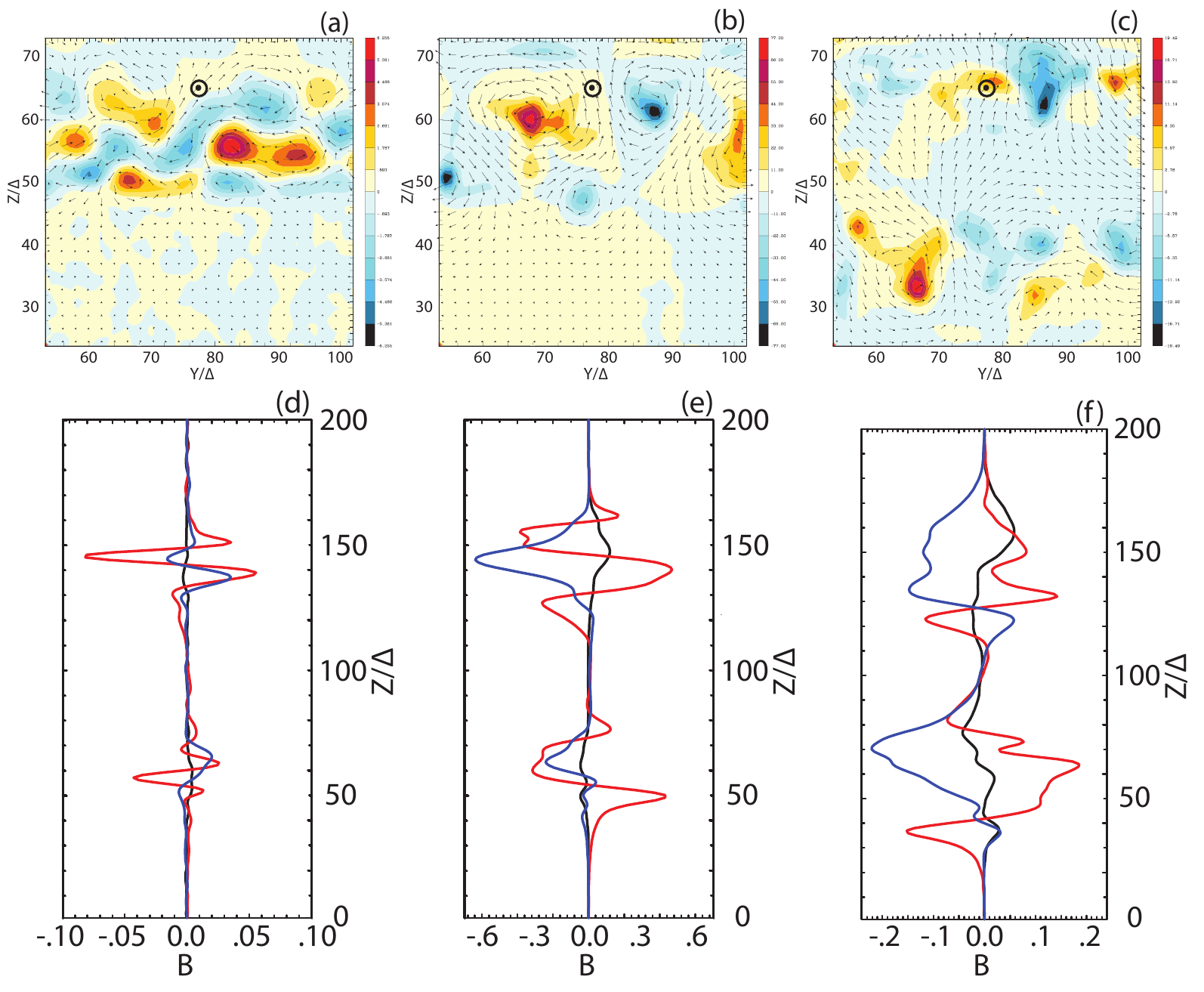}
\vspace{-0.3cm}
\caption{\footnotesize \baselineskip 11pt The  panels show the   $J_{\rm x}$  current structures   in a small square region (see Fig. \ref{Jxem}d)  for the $e^{\pm}$ cases with (a) $\gamma_{\rm jt} = 15$, (b) 
$\gamma_{\rm jt} = 5$, and  (c) $\gamma_{\rm jt} = 1.5$ in the $y - z$ plane at  $t = 300 \omega_{\rm pe}^{-1}$.  Note that panel (a) is the same as in Fig. \ref{Jxem}d. Corresponding 1D cuts along $z$ at $x = 500\Delta$ and $y = 100\Delta$ in the panels in the bottom row show $B_{\rm x}$ (black), $B_{\rm y}$ (red), and $B_{\rm z}$ (blue)  for the three Lorentz factor $e^{\pm}$ cases. The maximum and minimum current amplitude is $J_{\rm x} \sim$ (a) $\pm 6.26$, (c) $\pm 77.0$ (e) $\pm 19.49$.
\label{JxBem}}
\vspace{-0.10cm}
\end{figure}

Figure \ref{JxBem} shows how the $J_{\rm x}$ current  structure in a small $y - z$ plane square around the velocity shear surface (see Figure \ref{Jxem} for location) and the magnetic field strength and position relative to the shear surfaces along 1D cuts in the $z$ direction change as a function of the Lorentz factor for the $e^{\pm}$ cases at time $t = 300 \omega_{\rm pe}^{-1}$. 
{\kn Not presented here are the  $e^-$- $p^+$ cases as they show very little change in the amplitude or the width of the amplified magnetic field region as a function of the Lorentz factor. {\kn This result may indicate a difference from the theory and counter-streaming simulation results (Grismayer et al.\ 2013a,b; Alves et al.\ 2014) in which amplitude and width scaled with $\sqrt{\gamma_0}$ or just that our higher Lorentz factor $e^-$- $p^+$ cases have not reached saturation.} In the $e^{\pm}$ cases we see that the currents are located on the jet side of the shear surface for Lorentz factors $\gamma_{\rm jt} =$ (a) 15 and (b) 5, but are located on both sides of the shear surface for (c) $\gamma_{\rm jt} = 1.5$.  The maximum and minimum current density amplitudes are  (a) $\pm 6.26$, (b) $\pm77.0$, and (c)  $\pm19.5$, and the maximum magnetic field strength is smaller by about an order of magnitude for the (a) $\gamma_{\rm jt} = 15$  case and  smaller by about a factor of  3 for the (c) $\gamma_{\rm jt} = 1.5$ case compared to the (b) $\gamma_{\rm jt} = 5$ case.  Here we do find an  increase in the maximum field strength from the $\gamma_{\rm jt} = 1.5$ case to the $\gamma_{\rm jt} = 5$ case as suggested by the theory and counter-streaming results but with a decrease in the total shear layer width instead of the expected increase in shear layer width.  The $\gamma_{\rm jt} = 15$ simulation is not near saturation so cannot be directly compared to the lower Lorentz factor cases.  However, it is clear that  the increased inertia of the more relativistically moving jet plasma inhibits motion of jet electrons across the shear surface and affects the shear structure significantly compared to counter-streaming simulations in which both plasmas have the same inertia.}

{\kn Temporal development of the total magnetic field energy shown in Figure \ref{EMevol} shows that the still slowly growing magnetic field energy for the (b) $\gamma_{\rm jt} = 5$ case is comparable to the (c) $\gamma_{\rm jt} = 1.5$ case at time $t = 300 \omega_{\rm pe}^{-1}$ and should become greater at later times.  Figure \ref{EMevol}  also shows a more rapid growth of the total magnetic field energy for the (a) $\gamma = 15$ case at this time.  These results suggest that the differences in the current structure and the magnetic field strength and location may indicate a temporal development attributable to growth rate differences in addition to inertial differences.  In fact, for the case with $\gamma_{\rm jt} = 1.5$ at  $t = 200 \omega_{\rm pe}^{-1}$ the current filaments with maximum and minimum values $\pm 42.6$ are located nearer to the velocity shear than at the later time of $t = 300 \omega_{\rm pe}^{-1}$ shown in Figure \ref{JxBem}c where the maximum and minimum values are  $\pm 19.5$. 
Thus, the differences seen in Figure \ref{JxBem} from high to low Lorentz factors may provide an indication of the temporal development of the current structure from fewer ($\gamma_{\rm jt} = 15$) to more ($\gamma_{\rm jt} = 1.5$) e-folding times.} 

\vspace{-0.75cm}
\section{Microphysics at the Velocity Shear Surface}
\vspace{-0.25cm}

\subsection{3D Structure}
\vspace{-0.25cm}

\begin{figure}[th!]
\vspace{-0.5cm}
\epsscale{0.94}
\plotone{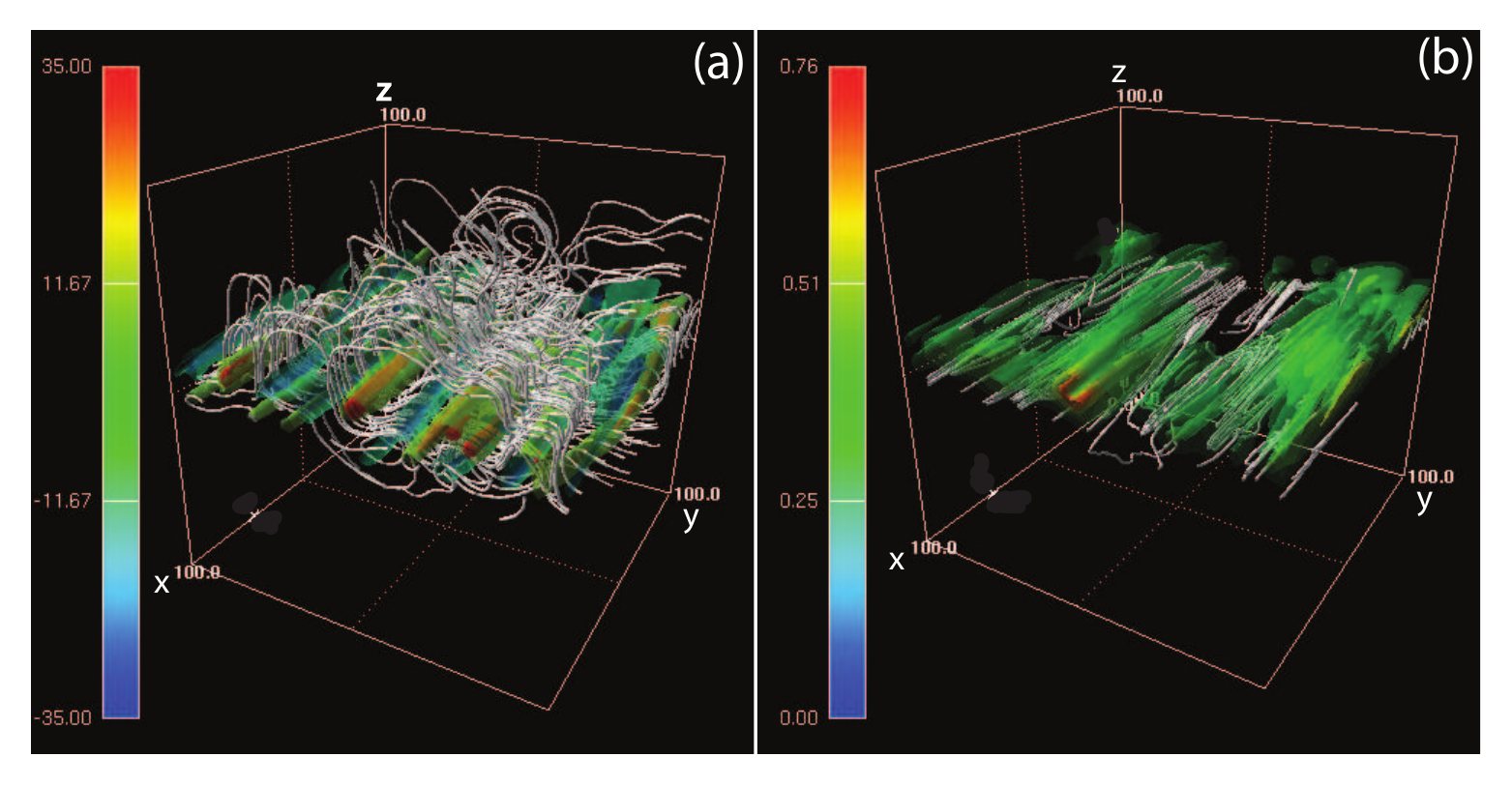}
\vspace{-0.3cm}
\caption{\footnotesize \baselineskip 11pt Current filament and magnetic field structure at the velocity shear surface displayed as (a) an isosurface of the $x$ component of current density with white magnetic field lines and (b) as an isosurface of the total magnetic field energy with white current stream lines  for  the $e^{\pm}$ case with $\gamma_{\rm jt} =  5$ at   $t = 250 \omega_{\rm pe}^{-1}$.  The volume $ 0 \le x/\Delta, y/\Delta, z/\Delta,  \le 100$ is displayed. 
\label{3Disos}} 
\vspace{-0.1cm}
\end{figure}

Figure \ref{3Disos} provides a 3D display of the currents and magnetic fields at the velocity shear surface for the $e^{\pm}$ case with $\gamma_{\rm jt} =$  5 at  $t = 250 \omega_{\rm pe}^{-1}$.  The 3D display reveals current filaments with length
in the $x$ direction (see Figure \ref{JXxz}d) much longer than the spacing in the $y - z$ plane (see Figure \ref{JxBem}b).  Strong positive (red) and negative (blue) current filaments wrapped by magnetic field lines seen in 3D are seen in a 2D slice shown previously in Figure\ \ref{JxBem}b (albeit here at an earlier time) in the  $x$ component of the current density (positive (red) and negative (blue)) and magnetic field (arrows) in the $y-z$ plane. The positive and negative current filaments seen in 2D are now revealed to twist around each other with the longitudinal wavelength $\lambda^{obs} \sim 100 \Delta$ seen in Figure \ref{JXxz}d and Figure \ref{By1dZ}d.  The total magnetic energy isosurface shows a concentration of the magnetic field around the current filaments. 

\begin{figure}[h!]
\vspace{-0.3cm}
\epsscale{0.85}
\plotone{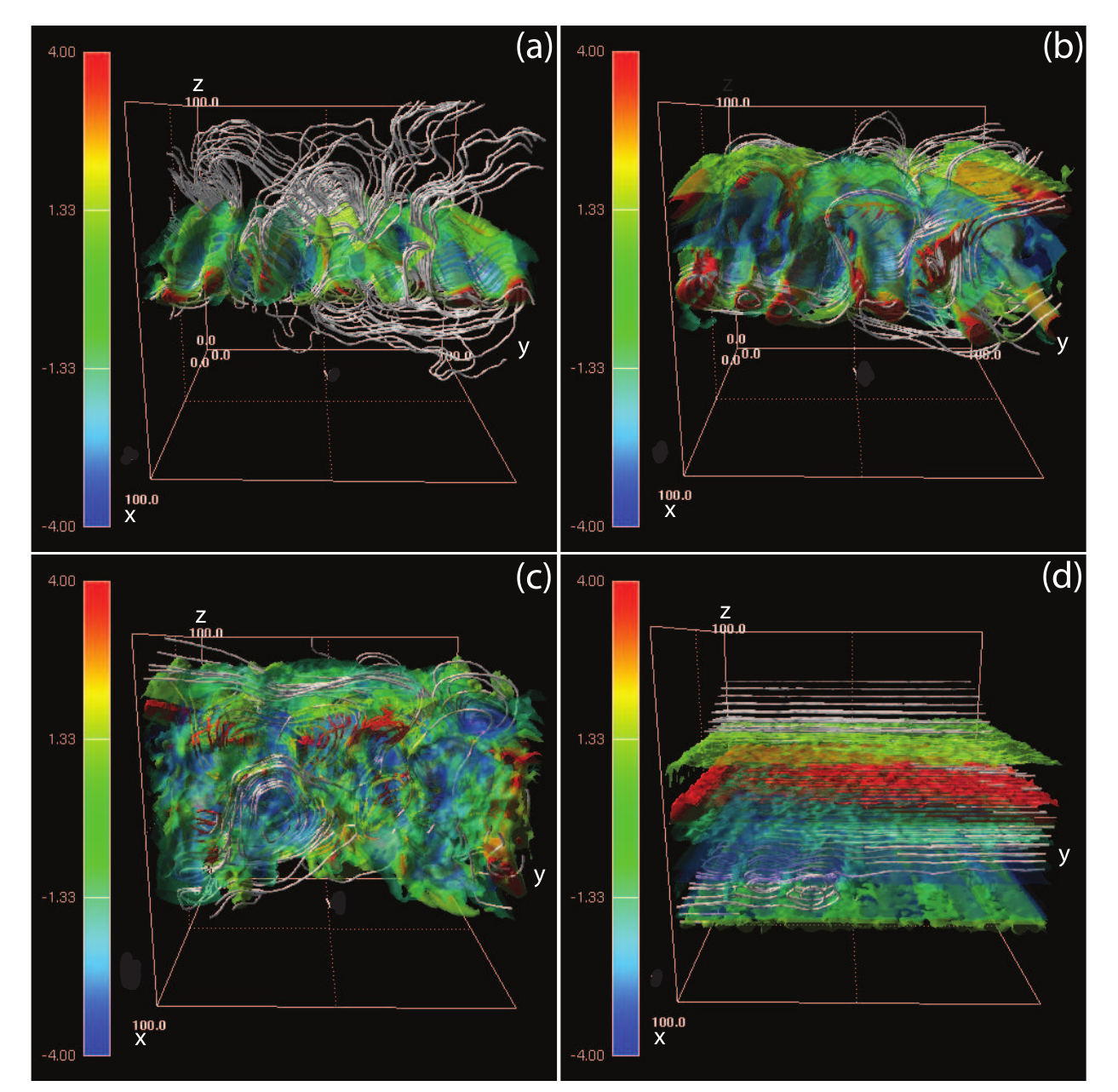}
\vspace{-0.3cm}
\caption{\footnotesize \baselineskip 11pt Currents and magnetic fields at the velocity shear surface displayed as isosurfaces of the $x$ component of current density with white magnetic field lines for $e^{\pm}$ cases with (panel a) $\gamma_{\rm jt} =  15$, (panel b)  $\gamma_{\rm jt} =  5$, and (panel c)  $\gamma_{\rm jt} =  1.5$ at $t = 300 \omega_{\rm pe}^{-1}$. These 3D panels correspond to Figure \ref{JxBem} panels a, b, and c, respectively.  Panel (d) shows the currents and magnetic fields for the $e^-$- $p^+$ case with  $\gamma_{\rm jt} =  5$ at $t = 300 \omega_{\rm pe}^{-1}$. The volume $ 0 \le x/\Delta, y/\Delta, z/\Delta,  \le 100$ is displayed. 
\label{FigT}}
\vspace{-0.1cm}
\end{figure}

Figure \ref{FigT} provides a 3D display of the currents and magnetic fields at the velocity shear surface 
for the $e^{\pm}$ cases  at  the same time, $t = 300 \omega_{\rm pe}^{-1}$, as the 2D slices shown in Figure \ref{JxBem}, and also shows the current and magnetic field structure at the velocity shear surface for the  $e^-$- $p^+$ case with  $\gamma_{\rm jt} =  5$.  For the $e^{\pm}$ cases, as indicated by the 2D slices, the 3D structure shows 
a current carrying region that thickens as the Lorentz factor decreases and at low Lorentz factor appears on both sides of the velocity shear surface.  The 3D structure suggests a single layer of current filaments at $\gamma_{\rm jt} =  15$ that broadens to a dual layer of current filaments at $\gamma_{\rm jt} =  5$. At $\gamma_{\rm jt} =  1.5$ current filaments on both sides of the velocity shear layer are much less well defined.  A comparison between the $\gamma_{\rm jt} =  5$ $e^{\pm}$ (Fig. \ref{FigT}b) and $e^-$- $p^+$ (Fig. \ref{FigT}d) cases shows the very different current and magnetic field structures at the velocity shear surface.  For the $e^-$- $p^+$case, the magnetic field is very uniform and largely confined to the velocity shear surface just below a strong positive (red) current sheet on the jet side. Outside the velocity shear surface we see a weaker negative (blue/green) current sheet, and further outside a filamented weak negative (blue/green) current region. 

The  structures shown in Figures \ref{3Disos} and \ref{FigT} are similar to those produced by the filamentation (Weibel-like) instability, associated with interpenetrating plasmas (see eq.(18)).  We note that the change in structure from closely spaced current filaments of smaller diameter in a narrower region in the $\gamma_{\rm jt} = 15$ case to the merged larger-diameter and less closely spaced filaments in a broader region in the $\gamma_{\rm jt} = 5$ case shown in Figures \ref{JxBem} and \ref{FigT} is like the expected temporal development for the filamentation instability as the number of e-folding times increases. Since  in our simulations the $\gamma_{\rm jt} = 5~\&~15$ cases have not reached saturation, one cannot say with certainty that the instability will not ultimately develop the structures seen in the saturated $\gamma_{\rm jt} = 1.5$ case, in which the current filaments are  probably fully developed by $t = 300 \omega_{\rm pe}^{-1}$.  However, it seems more likely that the lack of significant current structure on the outside of the velocity shear surface in the two higher Lorentz factor $e^{\pm}$ cases is a direct result of the increased inertia of the relativistically moving plasma.

\vspace{-0.75cm}
\subsection{Particle Motion}
\vspace{-0.20cm}

The observed 2D and 3D structures indicate the development of longitudinal (electrostatic two-stream) and transverse (Weibel-like current filamentation, e.g., Nishikawa et al.\ 2005, 2006, 2009) plasma instabilities  usually associated with interpenetrating relativistic plasmas.  
\begin{figure}[h!]
\vspace{-0.4cm}
\epsscale{0.97}
\plotone{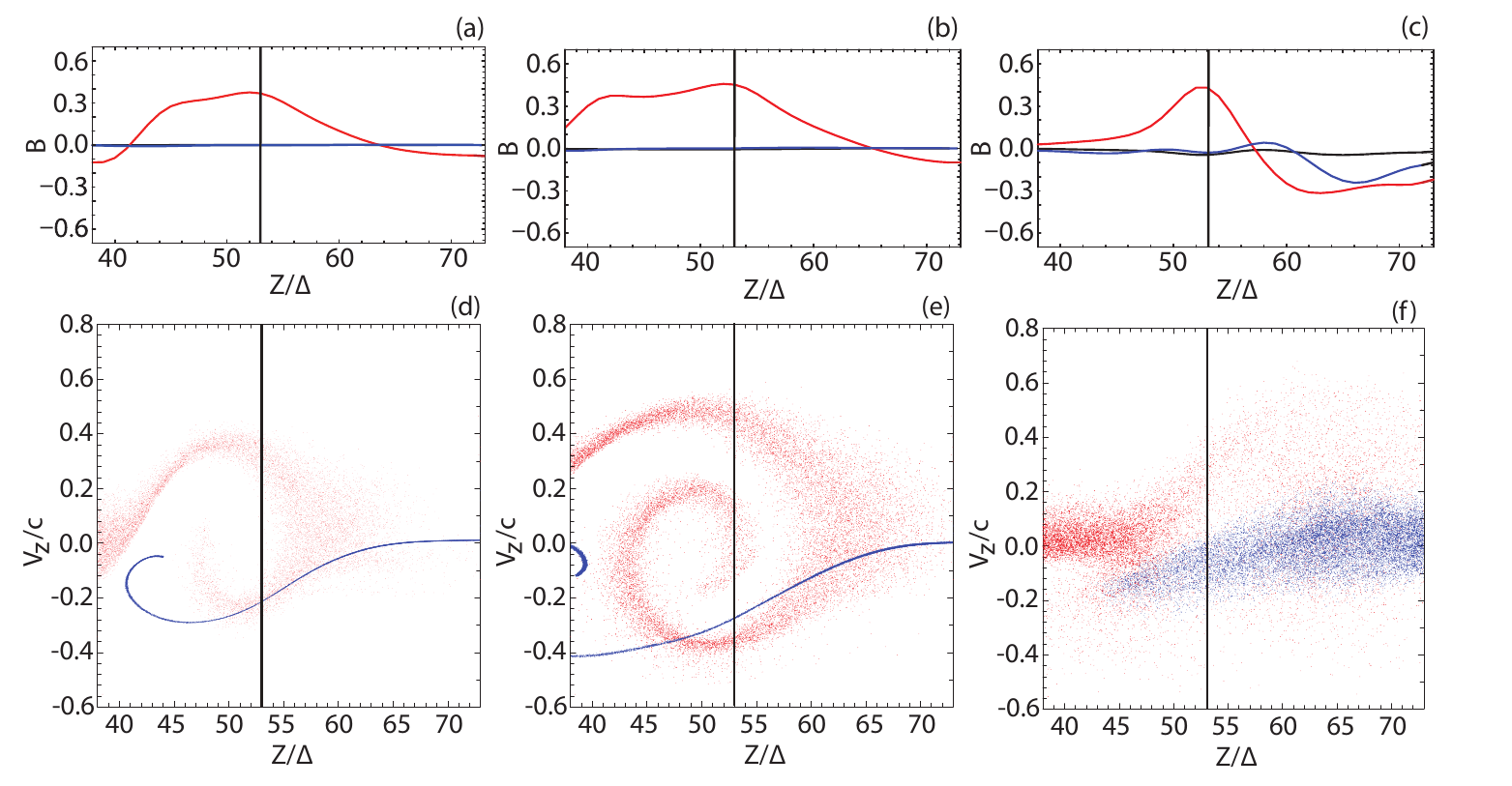}
\vspace{-0.3cm}
\caption{\footnotesize \baselineskip 11.0pt  The top row of panels shows the magnetic field components  $B_{\rm x}$ (black),  $B_{\rm y}$ (red), and $B_{\rm z}$ (blue)  at $x = 500\Delta$ and $y = 100\Delta$ for the $\gamma_{\rm jt} = 5$ $e^-$- $p^+$ case at (a) $t = 225 \omega_{\rm pe}^{-1}$ and (b) $t = 300 \omega_{\rm pe}^{-1}$, and (c) for the $e^{\pm}$ case at  $t = 300 \omega_{\rm pe}^{-1}$. Each panel in the bottom row corresponds to the top panel immediately above and provides a particle $v_{\rm z}-z$ phase-space plot near the velocity shear surface. In all panels the location of the velocity shear surface is indicated by the vertical line at $z /\Delta = 53$ initially with ambient (red) electrons to the left and jet (blue) electrons to the right of the shear surface.  
\label{phasBy}}
\vspace{-0.2cm}
\end{figure}
Figure \ref{phasBy} shows the magnetic field components and the associated phase-space plots of electron motions in the $z$ direction perpendicular to the velocity shear surface. These motions produce the mixing required to trigger interpenetrating plasma instabilities. Note that in $e^-$- $p^+$ cases,  mixing is almost completely associated with the electrons and in $e^{\pm}$ cases both electrons and positrons participate in the mixing. 

The $e^-$- $p^+$ $\gamma_{\rm jt} = 5$ case at  two different times illustrates the development of both the dominant $B_{\rm y}$ component of the magnetic field and the  plasma mixing process.  The magnetic field is initially strongest at the shear surface. The magnetic field strengthens and more deeply penetrates the ambient plasma with time.  Slightly deeper penetration into the jet plasma also occurs with time. Ambient electrons (red dots) with $v_z > 0$ are moving towards and into the jet, and become more heated and penetrate deeper into the jet plasma with time.  Relatively cold (note a very small thermal spread) jet electrons (blue dots) with $v_z < 0$ are moving outwards and into the ambient plasma. These jet electrons, while remaining cold, penetrate deeper into the ambient plasma with time. At the simulation times presented, the electrons are mixed in space but not yet in velocity. Due to the relatively uniform DC magnetic field generated in the $x$ and $y$ directions, the phase-space plot shows a regular structure. 

In the $e^{\pm}$ $\gamma_{\rm jt} = 5$ case, the dominant $B_{\rm y}$ component of the magnetic field is strongest at the shear surface and more deeply penetrates the jet plasma than the ambient plasma. Note the very different location of the magnetic field in this case versus the $e^-$- $p^+$ case at the same simulation time. The electrons are less mixed spatially on the ambient side of the shear surface but with much more heating of the jet electrons than in the $e^-$- $p^+$ case. Here most of the action resides on the jet side of the shear surface where the filamentation instability dominates the dynamics.  Both ambient and jet electrons are accelerated in the strong AC magnetic and electric fields associated with the filamentation instability. Ambient electrons are more strongly heated than the jet electrons, but now there is a significant velocity mixing of the jet and ambient electrons and the ambient electrons penetrate into the jet more deeply than in the $e^-$- $p^+$ case.

\begin{figure}[h!]
\vspace{-0.4cm}
\epsscale{0.9}
\plotone{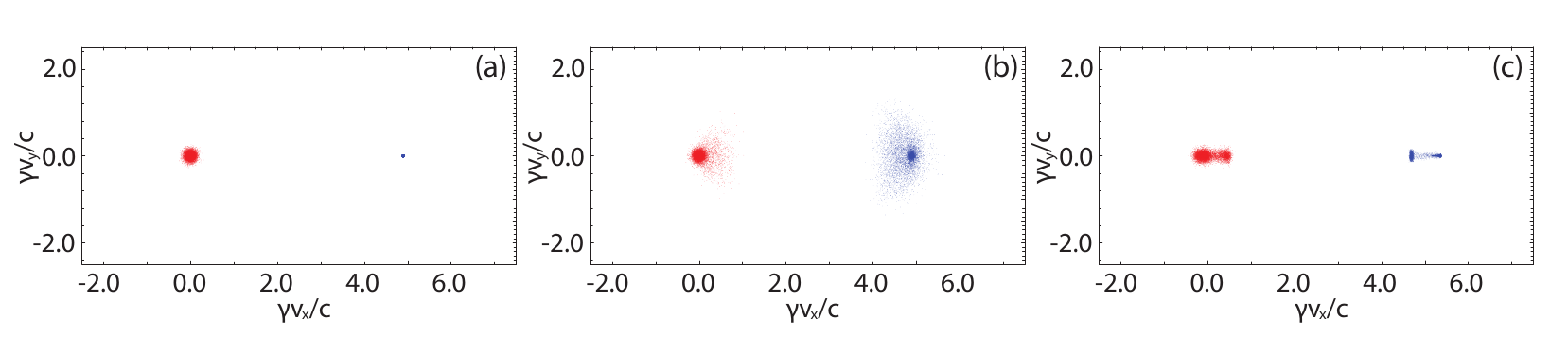}
\vspace{-0.3cm}
\caption{\footnotesize \baselineskip 11.0pt  The panels show a particle   $v_{\rm x}-v_{\rm y}$ plot for the $\gamma_{\rm jt} = 5$  simulation (a) initial conditions,  and (b) for the $e^{\pm}$ and (c)  $e^-$- $p^+$ cases at  $t = 300 \omega_{\rm pe}^{-1}$.  Jet (blue) electrons and  ambient (red) electrons are chosen randomly in the region  $5 < z/\Delta < 101$  near the velocity shear.  
\label{phasvx-vy}}
\vspace{-0.2cm}
\end{figure}

Just as Fig. \ref{phasBy} shows that the particle behavior near the velocity shear for the $e^{\pm}$ and  $e^-$- $p^+$  cases is significantly different,  Figure \ref{phasvx-vy} shows that electron acceleration at the velocity shear also is significantly different in $e^{\pm}$ and  $e^-$- $p^+$  cases. 
The  $v_{\rm x}-v_{\rm y}$  velocity component plots show that  the electrons are accelerated in parallel and perpendicular directions (Figure \ref{phasvx-vy}b) in the $e^{\pm}$ case and the electrons are accelerated only in the parallel direction (Figure \ref{phasvx-vy}c)  in the $e^-$- $p^+$ case.

Similar analysis for the two other jet Lorentz factors (not shown) shows that $e^-$- $p^+$ cases with different Lorentz factor show the same parallel electron acceleration.   On the other hand, there are some modest differences in the $e^{\pm}$ cases for the two other jet Lorentz factors studied.  These differences occur because in the $\gamma_{\rm jt} = 1.5$ case the ambient and jet electrons are strongly mixed across the velocity shear surface, but in the $\gamma_{\rm jt} = 15$ case, the ambient electrons are strongly mixed into the jet region, but jet electrons are only weakly mixed into the ambient plasma. {\kn This velocity and phase-space result is  also revealed in Figures \ref{JxBem} and
\ref{FigT} that show a development of current filament structure on both sides of the velocity shear surface for the $\gamma_{\rm jt} = 1.5$ case but only on one side of the velocity shear surface for the $\gamma_{\rm jt} = 15$ case. In the $\gamma_{\rm jt} = 1.5$ case the mixing of jet and ambient electrons on both sides of the velocity shear is accompanied by mixing in $v_{\rm x} - v_{\rm y}$ and acceleration in both parallel and perpendicular directions like the $\gamma_{\rm jt} = 5$ case. In the $\gamma_{\rm jt} = 15$ case ambient and jet electrons are accelerated
mainly in the perpendicular direction.}

{\kn Our simulations do not follow the kKHI significantly past the saturation phase (see Section 3.1) and are thus too short to allow for significant electron acceleration in the self-generated fields of the shear flow instabilities. In fact, the strongest acceleration should occur after the growth of the kKHI fully saturates and electrons can probe the long-lived turbulent electric fields in the shear region. This has been demonstrated by Alves et al.\ (2014) for the electron-proton plasma, who discuss the acceleration process and estimate  the maximum energy gain of an electron interacting in the electric field to be $\Delta \mathcal{E} _{\max}  \propto mc^{2}\gamma_{0}^{4}$. 
Acceleration to super-thermal energies is thus possible for shear flows with relativistic Lorentz factors as shown by the PIC simulations in Alves et al.\ (2014). In our simulations we trace the initial stages of electron acceleration. We note that the acceleration in the parallel direction observed in our $e^-$- $p^+$ simulations resemble a process of electron surfing on the electric field structures in the shear region that provides acceleration mostly in the direction of the bulk plasma flow, as indicated in Alves et al. (2014).}

\vspace{-0.75cm}
\section{Summary}   
\vspace{-0.2cm}

We have presented 3D PIC simulations of the kinetic Kelvin-Helmholtz instability for both electron-positron and electron-proton plasmas. The processes studied here are of importance to  the jets from AGNs and GRBs that are expected to have velocity shears between faster and slower moving plasmas both within the jet and at the jet external medium interface. In our simulations we have studied large velocity shears with relative Lorentz factors of 1.5, 5, and 15. The simulations are performed in the rest frame of an ambient plasma sheath, and an appropriate Lorentz transformation of the results will extend the analysis to an ambient plasma sheath of arbitrary speed.

Our work goes beyond the scope of earlier 2D simulations performed by Liang et al.\ (2013a,b) in either the shear momentum parallel plane ($x - y$ referred to as $P$) or the transverse plane ($y - z$ referred to as $T$). The full three-dimensional effects that we find here are not found in their simulations. We show that the kinetic Kelvin-Helmholtz instability depends on the composition of  the plasma and the jet Lorentz factor. The electron-proton cases generate a DC magnetic field in the shear plane, perpendicular to the relative velocity ($B_{\rm y}$ with $E_{\rm z}$), while on the contrary, the electron-positron cases generate AC electric and magnetic fields.  In the electron-positron cases current filaments are generated similar to those found associated with the filamentation (Weibel-like) instability. In the simulations, initial growth appears in the $E_{\rm z}$ electric field component perpendicular to the velocity shear surface.  This growth is followed by growth of the $B_{\rm y}$ magnetic field component in the velocity shear plane transverse to the flow direction in the electron-proton cases. For the electron-positron cases, growth is seen in both $B_{\rm y}$ and $B_{\rm z}$ magnetic field components as current filaments dominate the structure near the velocity shear surface. In all cases, fluctuation structure along the jet is much longer than transverse fluctuation structure. For electron-proton cases interaction and magnetic fields do not extend far from the initial velocity shear surface. For the electron-positron cases, interaction and magnetic fields extend farther from the initial velocity shear surface although they extend mostly into the jet side for higher jet Lorentz factor. 

The velocity shear behavior of the magnetic fields should have consequences for the appearance of jets in very-high-resolution radio imaging. For a simple cylindrical geometry velocity shear case, an electron-proton jet would primarily build magnetic field in the toroidal direction at the velocity shear surface.  The magnetic field would be quasi-parallel to the line of sight at the limbs of the jet for typical aspect angles $\theta\approx \gamma_\mathrm{jet}^{-1}$.   In contrast, a pair-plasma jet would generate sizable radial field components that are only about a factor of 2 weaker than the toroidal field. 

The strong electric and magnetic fields in the velocity shear zone will also be conducive to particle acceleration. Our simulations are too short for definitive statements on the efficacy of the process and the resulting spectra. Also, the organization of the field in compact regions will complicate the interpretation of emission spectra, and a spatially-resolved treatment of particle acceleration and transport would be mandatory for a realistic assessment, which is beyond the scope of the present paper. Relativistic electrons, for example, will suffer little synchrotron energy loss outside of the thin layer of strong magnetic field. Thus synchrotron emissivity will be dominated by the shear layer, and in general emissivity will depend on how efficiently electrons can flow in and out of the shear layer  and be accelerated in the regions of strong magnetic field. An immediate consequence for radiation modeling is that the energy loss time of electrons cannot be calculated with the same mean magnetic field that is used to compute emission spectra, because the former includes the volume filling factor of the strong-field regions.

{\kn We have extended the stability analysis presented in Gruzinov (2008), Alves et al.\ (2012) and Alves et al.\ (2014) to core-sheath electron-proton plasma flows allowing for different jet core and ambient sheath electron densities $n_{\rm jt}$ and $n_{\rm am}$, respectively, and jet core and ambient sheath electron velocities $v_{\rm jt}$ and $v_{\rm am}$, respectively. In this analysis the protons are considered to be infinitely massive and free-streaming,  whereas the electron fluid quantities and fields are linearly perturbed. Not unexpectedly we find a smaller temporal growth rate for larger Lorentz factors, although in the simulations the growth rate does not appear to decrease as rapidly with Lorentz factor as the maximum growth rate, $\omega^* \propto \gamma_{\rm jt}^{-1}$ (timescales $t \propto \gamma_{\rm jt}$), obtained from the longitudinal dispersion relation.  It is likely that the growth of the transverse structure seen in the 3D simulations, which likely grows on timescales $t \propto \gamma_{\rm jt}^{1/2}$, is responsible for the difference. Fluctuation wavelengths along the flow direction seen in the two lower Lorentz factor simulations are of the order of the predicted fastest growing wavelengths for both electron-proton and electron-positron plasmas suggesting that the dispersion relation applies approximately even for a non-free streaming equal mass positively charged particle.  On the other hand, the rate of growth and non-linear structure is very different for electron-proton and electron-positron plasmas.}

\acknowledgments

This work is supported by NSF AST-0908010, and AST-0908040, NASA-NNG05GK73G,
NNX07AJ88G, NNX08AG83G, NNX08AL39G,  NNX09AD16G,  NNX12AH06G,
NNX13AP-21G, and NNX13AP14G. The work of JN has been supported by the Polish National Science Centre through projects DEC-2011/01/B/ST9/03183 and DEC-2012/04/A/ST9/00083.
YM is supported by the Ministry of Science and Technology of Taiwan
under the grant NSC 100-2112-M-007-022-MY3. M.P. acknowledges support through grant PO 1508/1-2 of the Deutsche
Forschungsgemeinschaft. Simulations were performed at the Columbia and Pleiades facilities at the
NASA Advanced Supercomputing (NAS) and Kraken and Nautilus at The National Institute for Computational Sciences (NICS) which is supported by the NSF. This research was started during the program ``Chirps, Mergers and Explosions:  The Final Moments of Coalescing Compact Binaries'' at the Kavli Institute for Theoretical Physics which is supported by  the National Science Foundation under Grant No. PHY05-51164. The first result of the kKHI with mi/me = 1 was obtained during the Summer Aspen 
workshop ``Astrophysical Mechanisms of Particle Acceleration and Escape from the Accelerators''
held at the Aspen Center for Physics (September 1 - 15, 2013).

\end{document}